\documentclass{aa}  

\usepackage{lineno}
\renewcommand{\linenumbers}{}

\usepackage{graphicx}
\usepackage{txfonts}
\usepackage{lipsum}
\usepackage{subcaption}         
                                
\usepackage{lscape}             
                                
\usepackage{placeins}

\usepackage{hyperref}

\usepackage{threeparttable}
\usepackage{xcolor}
\begin{document}

   \title{Investigating disc-corona interaction in axisymmetric accretion disc models}

   \subtitle{}

   \author{Lixin Zhang\inst{1}
        \and Li Xue\inst{1}
        \and Jingyi Luo\inst{1}
        \and Chengzhi Li\inst{1}
        }

   \institute{Department of Astronomy, Xiamen University, 361005, Fujian, China\\
   \email{lixue@xmu.edu.cn}}

   \date{Received September 30, 20XX}
 
  \abstract
   {The interaction between the accretion disc and its corona plays a critical role in the energy balance and emission mechanisms in astrophysical systems such as active galactic nuclei and X-ray binaries. However, the detailed physics of disc-corona interactions, including the mechanisms driving disc evaporation and the impact of the accretion rate and viscosity, remain poorly understood.}
   {Our study aims to extend the well-known disc evaporation model to investigate the disc-corona interaction in a 2D axisymmetric, time-dependent hydrodynamic model, focusing on the effects of the viscosity and accretion rate, and their influence on disc evaporation, luminosity, and corona formation.} 
   {We developed a hydrodynamic model consisting of a thin accretion disc, a corona, and a vacuum region. Our model was implemented in Athena++, with the gas-vacuum interface tracking algorithm to handle the vacuum regions. We performed simulations incorporating turbulent viscosity, thermal conduction, bremsstrahlung cooling, and artificial disc cooling, starting from an adiabatic state to explore the disc-corona interaction.}
   {We demonstrate the presence of acoustic shock heating. We find that viscosity dominates the intensity of disc evaporation, that the accretion rate primarily determines the disc truncation radius and the disc luminosity, and that there may be a positive correlation between the corona luminosity and the evaporation intensity. We find the warm gas required by the warm corona model. We also compare our results with observations and simulations, and estimate the $y$ parameters to explore the potential effects of Compton cooling as well as the potential effects of the warm corona.}
   {}


   \maketitle

\section{Introduction}\label{sec:intro}

The state transition is an essential observational phenomenon in X-ray binaries (XRBs), including two subtypes of low-mass systems (neutron star X-ray binaries, NSXBs) and high-mass systems (black hole X-ray binaries, BHXBs). This phenomenon has been widely studied in the community of high-energy astrophysics for more than 30 years \citep{Tanaka1996, Remillard2006, Done2007, Fender2012, Zhang2013, Liu2022}. The first state transition was observed in an NSXB by \cite{Mitsuda1989}, and later in a BHXB by \cite{Ebisawa1994}. \cite{Esin1997} produced the first modelling spectrum for a BHXB with a hybrid model composed of an inner advection-dominated accretion flow \citep[ADAF,][]{Narayan1994} and an outer Shakura–Sunyaev accretion disc \citep[SSD,][]{Shakura1973}. \cite{Rozanska2000} presented an analytical algebraic model of the disc-corona accretion flow around the black hole, which was the first theoretical work of disc evaporation for supermassive active galactic nuclei (AGNs). Following closely behind, \cite{Meyer2000} developed a vertical equilibrium disc evaporation model (DEM) to explain state transitions in BHXBs, which was adapted from their previous model for white dwarf accretion \citep{Meyer1994}. \cite{Mayer2007} proposed the first time-dependent one-dimensional (1D) radial model, which incorporates both the corona and the accretion disc, with mass exchange between them occurring through physical processes of evaporation and condensation. Driven by advancements in computational technology, hydrodynamic (HD) and magnetohydrodynamic (MHD) simulations have been widely applied to accretion studies, which capture the physical processes of disc-corona interactions \citep[e.g.][]{Wu2016,Jiang2019,Nemmen2024}.

Since the seminal work of \cite{Meyer2000}, the DEM has been significantly promoted to explain many observational phenomena from stellar-mass XRBs to supermassive AGNs \citep[e.g.][]{Meyer2002, Liu2002, Qian2007, Taam2008, Qiao2013, Liu2015, Cheng2020, Cho2022}. After years of development, DEM already contains several important physical mechanisms, including turbulent viscosity, magnetic reconnection heating, thermal conduction, bremsstrahlung, and inverse Compton radiation cooling \citep{Liu2022}. These mechanisms can be divided into heating evaporation mechanisms (including heating from turbulent viscosity, magnetic reconnection, and thermal conduction on cold gas) and cooling condensation mechanisms (including cooling from bremsstrahlung radiation, inverse Compton radiation, and thermal conduction on hot gas), based on their effects on the gas state. Therefore, in DEM, the explanation of state transition in XRBs is achieved by simultaneously exploring the equilibrium between these mechanisms. However, it is incomplete as DEM is a vertical 1D model that reaches the elaborate equilibrium only locally in the vertical direction at a specific radius of the accretion disc. Further improvements to DEM should at least consider the influence of two-dimensional (2D) axisymmetric geometric effects, which is the fundamental assumption adopted in most existing simulations of accretion discs. 

We notice that there have already been two studies focusing on the 2D simulation of hot flow cooling condensation by various radiation cooling processes \citep{Wu2016, Nemmen2024}. By contrast, studies on the heating evaporation of thin discs are currently absent in the community of astrophysical simulations (though a series of works on the time evolution of accretion discs have produced some sort of coronae, e.g. the series of papers starting from \cite{Jiang2019}). Thus, beginning with this paper, we are dedicated to gradually establishing a time-dependent axisymmetric DEM (poloidal 2D) based on the well-developed steady DEM. As the first step, we only focus on simulating the evaporation of thin discs under Newtonian gravity in order to address the numerical technical difficulties that need to be overcome during the initial establishment of this model (see Sect.~\ref{Sec:VRS}).

This paper is organised as follows. In Sect.~\ref{Sec:Model}, we introduce the basic equations and relevant settings used in our simulations. Sect.~\ref{Sec:VRS} provides a detailed explanation of the vacuum Riemann solver (VRS) and the gas-vacuum interface (GVI) tracking algorithm, both of which are critical components of our simulations. In Sect.~\ref{Sec:Results}, we present and discuss the results of our numerical simulations. Finally, in Sect.~\ref{Sec:Conclusion}, we summarise our findings and discuss potential further improvements.

\section{The model}\label{Sec:Model}
\subsection{Basic equations}
We considered a HD accretion model comprising a cold thin disc (cold and dense flow) and a concomitant hot extended corona (hot and tenuous flow). The system is governed by the following basic equations:
\begin{gather}
	\frac{\partial\rho}{\partial t} + \nabla\cdot\left(\rho\mathbf{v}\right) = 0, \\
	\frac{\partial\left(\rho\mathbf{v}\right)}{\partial t} + \nabla\cdot\left(\rho\mathbf{v}\mathbf{v}+p\mathbf{I}-\mathbf{T}\right) = -\rho\nabla\Phi, \label{eq_momentum} \\
	\frac{\partial e}{\partial t} + \nabla\cdot\left[\left(e+p\right)\mathbf{v}-\mathbf{v}\cdot\mathbf{T}+F_{\rm{c}}\right] = -\rho\mathbf{v}\cdot\nabla\Phi-q_{\rm{r}}, \label{eq_energy}
\end{gather}
where $t$, $\rho$, $e$, $p$, $\mathbf{v}$, $\Phi$, $\mathbf{T}$, $F_{\rm{c}}$, and $q_{\rm{r}}$ represent the time, density, total energy, pressure, velocity vector, Newtonian gravitational potential, viscous stress tensor, heat flux, and radiation cooling rate, respectively. We solved these equations in spherical polar co-ordinates, using 2D axisymmetric simulations with Athena++ \citep{Stone2008}. The velocity vector and total energy are expressed as
\begin{gather}
	\mathbf{v} = \left(v_r, v_{\theta}, v_{\phi}\right), \\
	e = \frac{p}{\gamma - 1} + \frac{1}{2} \rho \mathbf{v}^2, \label{eq_def}
\end{gather}
where $v_r$, $v_{\theta}$, and $v_{\phi}$ are the velocity components in the radial, poloidal, and toroidal directions, respectively, and $\gamma$ (set to $5/3$) is the adiabatic index for a non-relativistic ideal gas. Other quantities, such as $\mathbf{T}$, $F_{\rm{c}}$, and $q_{\rm{r}}$, are all defined in subsequent sections.

\subsection{Model settings}
This section outlines the physical (gravity, viscosity, thermal conduction, and radiation cooling) and numerical (computational domain, boundary, and initial conditions) settings used in the simulations.

\subsubsection{Gravity}
For gravity, we used the Newtonian point-mass gravitational module in Athena++, with the point mass located at the co-ordinate origin, which is applicable only in spherical polar co-ordinates ($r, \theta, \phi$). The black hole mass was set to $M = 10M_{\odot}$, and we applied Newtonian gravity for simplicity, although Athena++ also includes a general relativity module. This choice was made to focus on the accretion process in the initial study, without dealing with the complexities of relativistic gravity.

\subsubsection{Viscosity}
For viscosity, we adopted an anisotropic viscous stress tensor, $\mathbf{T}$, following \cite{Stone1999}, for which only the azimuthal shearing components are non-zero:
\begin{gather}
	T_{r\phi} = \rho \nu r \frac{\partial}{\partial r}\left(\frac{v_{\phi}}{r}\right), \\
	T_{\theta\phi} = \frac{\rho \nu \sin\theta}{r} \frac{\partial}{\partial \theta}\left(\frac{v_{\phi}}{\sin\theta}\right).
\end{gather}
This set-up approximates the magnetic stresses associated with MHD turbulence driven by the magneto-rotational instability (MRI) \citep{Balbus1991, Balbus1998}, and has been widely adopted in other HD simulations \citep[e.g.][]{Yuan2012, Wu2016, Nemmen2024}. We used the $\alpha$-viscosity prescription \citep{Shakura1973}, with the kinematic viscosity coefficient defined as
\begin{gather}
	\nu = \alpha \frac{c^2_{\rm{s}}}{\Omega_{\rm{K}}},
\end{gather}
where $\alpha$ is the viscosity parameter, $c_{\rm{s}} = \sqrt{p/\rho}$ is the sound speed, and $\Omega_{\rm{K}} = \sqrt{GM/r^3}$ is the Keplerian angular velocity.

Since the theory of MRI was proposed, the value of $\alpha$ has become an interesting issue investigated by many studies on accretion discs. \cite{King2007} summarised from observations that $\alpha$ typically lies in the range $0.1 < \alpha < 0.4$. However, many early MHD numerical simulations including local shearing box, global pseudo-Newtonian, or relativistic ones show that $\alpha<0.1$ \citep[e.g.][]{Hawley1995, Hawley2001, McKinney2012, Penna2013}. Recently, high-resolution global three-dimensional (3D) general relativistic MHD (GRMHD) simulations have shown the radial variation in $\alpha$, whose value varies from $0.001$ to $1$ \citep[see their Fig.~19 in ][]{Porth2019}. For simplicity, we ignored the radial variation in $\alpha$ noted by \cite{Penna2013} and \cite{Porth2019} to set it as a constant. In this work, we have adopted two typical values for the viscosity parameter, $\alpha = 0.3$ and $0.9$, which are in line with the DEM of \cite{Liu2002}. 

\subsubsection{Thermal conduction}
For thermal conduction, we followed \cite{Meyer2000} and used the Spitzer-Harm heat flux formula:
\begin{gather}
	F_{\rm{c}} = -\kappa_0 T^{5/2} \nabla T,
\end{gather}
where $\kappa_0 = 10^{-6}$ (in cgs units) and $T$ is the gas temperature. This formula is widely used in astrophysical models, though it predicts a physically unrealistic flux for hot and tenuous gases. Therefore, a saturated heat flux \citep{Luciani1983} is adopted:
\begin{gather}
	F_{\rm{s}} = \frac{1}{4} \frac{k^{3/2}}{m^{1/2}_{\rm{e}}} n_{\rm{e}} T^{3/2},
\end{gather}
where $n_{\rm{e}}$, $m_{\rm{e}}$, and $k$ are the electron number density, electron mass, and Boltzmann constant, respectively. In our simulations, the numerical heat flux was set to the minimum of $F_{\rm{c}}$ and $F_{\rm{s}}$.

It should be noted that we used the thermal conduction implementation in Athena++, which calculates $p/\rho$ instead of the exact temperature in evaluating conductive fluxes. This approximation is valid in the coronal region where radiation pressure can be ignored, but may overestimate the temperature in the optically thick disc flow. However, this overestimation has a minimal impact on thermal conduction in the disc gas, as the temperature gradient within the disc flow is relatively small.

\subsubsection{Radiation cooling}\label{subsubsec:rad_cooling}
For radiation cooling, we adopted different methods to treat the cooling in the coronal and disc flows. In the coronal flow, only bremsstrahlung cooling was considered, while synchrotron cooling and its Comptonisation, previously considered in condensation simulations \citep{Wu2016,Nemmen2024}, were neglected. Compton cooling caused by irradiation from the disc was also ignored, due to the complexity of implementing it in Athena++, which is beyond the scope of this paper. In the disc flow, we applied an artificial cooling model introduced by \cite{Noble2009}, which has proven to be robust and effective in relativistic disc simulations.

To simplify the calculation, we neglected the effect of radiation transfer, despite our model encompassing both optically thin and thick flows. To determine the cooling rate for each computational cell (CC), we used a simple strategy by selecting the lower value between bremsstrahlung cooling and artificial disc cooling:
\begin{gather}
q_{\rm{r}} = \min(q_{\rm{a}}, q_{\rm{b}}),
\end{gather}
where $q_{\rm{b}}$ represents the bremsstrahlung cooling rate, which is the sum of electron-ion and electron-electron cooling processes, calculated following \cite{Narayan1995}. $q_{\rm{a}}$ is the artificial disc cooling rate, defined as follows:
\begin{gather}
q_{\rm{a}} = \frac{p\Omega}{\gamma - 1} \left( Y - 1 + |Y - 1| \right)^w, \label{eq_artfcoolingrate} \\
\Omega \equiv \frac{v_{\phi}}{r}, \\
Y \equiv \frac{p}{\rho} \frac{\rho_{\rm{d}}}{p_{\rm{d}}}, \label{eq_Y}
\end{gather}
where $\Omega$ is the angular velocity, and $p_{\rm{d}}$ and $\rho_{\rm{d}}$ are the pressure and density of the SSD, which are obtained from the power-law formulas of \cite{Shakura1973} \citep[see also][]{Kato2008}. The term $Y - 1 + |Y - 1|$ in Eq.~(\ref{eq_artfcoolingrate}) acts as a switching function that enables rapid cooling on the orbital timescale or maintains the gas temperature close to the target SSD temperature. The exponent $w$ controls the growth of $q_{\rm{a}}$ when the gas temperature exceeds that of the SSD, and we adopted $w = 1/2$ as was suggested by \cite{Noble2009}. Gas with bremsstrahlung cooling was classified as coronal flow, while gas with artificial disc cooling was treated as disc flow. This strategy ensures a smooth transition between two different radiation cooling rates in the absence of consistent radiation transfer, while it can also select the proper radiation cooling rate to avoid unnecessary overcooling in the extreme cases of optically thin and thick flows.

For optically thick accretion discs, radiation pressure cannot be neglected, which is incorporated into our artificial disc cooling model. In Eq.~(\ref{eq_artfcoolingrate}), $p_{\rm{d}}$ includes both gas and radiation pressures (calculated from the disc temperature given by the power-law formulas of \cite{Kato2008}). The ratio $p_{\rm{d}} / \rho_{\rm{d}}$ represents the local internal energy of the SSD, and $Y$ in Eq.~(\ref{eq_Y}) quantifies the deviation between the internal energy of our time-dependent model and that of the SSD. When $q_{\rm{a}} < q_{\rm{b}}$, artificial disc cooling is activated, which constrains the pressure $p$ around $p_{\rm{d}}$, thereby incorporating radiation pressure. The effect of radiation pressure is then transmitted through the pressure gradient in the momentum equation (Eq.~\ref{eq_momentum}). If $q_{\rm{a}} > q_{\rm{b}}$, bremsstrahlung cooling is applied, where the temperature is determined solely by gas pressure, consistent with the coronal gas model that neglects radiation pressure. To prevent excessive cooling from stiff source terms, we imposed a floor on the internal energy (equivalent to the gas internal energy corresponding to a temperature of $10^4$ K).

\subsubsection{Computational domain and boundary conditions}
The computational domain is a sector-shaped area defined by four boundaries: the polar ($\theta=0$), equatorial ($\theta=\pi/2$), inner ($r=10r_{\rm{s}}$), and outer ($r=50r_{\rm{s}}$) boundaries, where $r_{\rm{s}} = \frac{2GM}{c^2}$ is the Schwarzschild radius. This domain is non-uniformly divided into 150 and 300 partitions in the radial and poloidal directions, respectively, resulting in a total of 45,000 CCs. The cell concentration can be adjusted by setting the size ratio between neighbouring cells to increase the spatial resolution near the inner and equatorial boundaries, ensuring computational precision in these critical regions. The highest resolution is achieved at the inner boundary, where the minimal radial width and equatorial vertical height of CCs are $\sim1.07\times 10^{-1}r_{\rm{s}}$ and $\sim7.50\times 10^{-4}r_{\rm{s}}$, respectively. This vertical resolution is sufficient to divide the typical half-thickness of the initial SSD into dozens of CCs at the inner boundary (for example, the half-thickness is $\sim4.57\times 10^{-2}r_{\rm{s}}$, which can be divided into $41$ CCs for the initial SSD of case D defined in Table~\ref{tab:DiffCases}). Symmetric boundary conditions (BCs) were applied on both the polar axis and equator. Unidirectional outflow BCs (which permit only the outflowing and prevent the inflowing) were first introduced in our previous work \citep{Xue2021}, and were applied here to both the inner boundary and most of the outer boundary. The remainder of the outer boundary, which corresponds to the SSD half-thickness near the equator, was set to an inflow boundary by fixing the initial SSD state at that radius. Therefore, the outer boundary is a combination of outflow and inflow BCs.

\subsubsection{Initial condition}\label{subsubsec:initialcondition}
In this study, we focus on the accretion surrounding a stellar mass black hole with a mass of $10M_{\odot}$. We ran four cases with different viscosities and accretion rates, as is listed in Table~\ref{tab:DiffCases}. These cases use typical parameters from the DEM \citep[see][]{Liu2002}. Due to the significant density variations in the disc evaporation process, we introduced a novel technique to incorporate the vacuum state into the classical finite volume method (FVM) simulation (see Sect.~\ref{Sec:VRS} for details). The initial condition in our simulations is simple: an SSD embedded in vacuum. The equatorial boundary is a geometric interface with no thickness, where the accretion gas is symmetrically distributed on both sides. We assume that physical quantities are uniformly distributed vertically within the SSD. Therefore, we only need to compute the quantities at the disc's central plane using the power-law formulas of SSD \citep{Shakura1973,Kato2008}, and then extend this layout vertically along the equatorial boundary up to the SSD scale height.

\renewcommand{\arraystretch}{1.5}  
\begin{table}[h!]
\begin{threeparttable}
\caption{Parameters and some results of four cases}                 
\label{tab:DiffCases}    
\centering                        
\begin{tabular}{c c c c c c}      
\hline\hline               
Case & A & B & C & D  \\        
\hline                     
$\alpha$ & 0.9 & 0.9 & 0.3 & 0.3 \\
$\dot{m}$ $(\dot{M}_{\rm{Edd}})$ \tnote{a} & 0.01 & 0.1 & 0.01 & 0.1 \\
$r_{\rm{tr}}$ ($r_{\rm{s}}$) \tnote{b} & 10.37 & 6.52 & 9.31 & 5.97 \\
$L_{\rm{disc}}$ ($L_{\rm{Edd}}$) \tnote{c} & 0.045 & 0.71 & 0.039 & 0.40 \\
$L_{\rm{disc}}/L_{\rm{coro}}$ \tnote{d} & 22.2 & 129.7 & 52.0 & 335.5 \\
$t_{\rm{R}}$ $(\times10^5r_{\rm{s}}/c)$ \tnote{e} & 3.09 & 2.63 & 3.00 & 3.24 \\
\hline                                 
\end{tabular}
\begin{tablenotes}\footnotesize
	\item[a] The accretion rate scaled by the Eddington accretion rate, $\dot{M}_{\rm{Edd}}$
	\item[b] The truncation radius
	\item[c] The disc luminosity scaled by the Eddington luminosity, $L_{\rm{Edd}}$
	\item[d] The ratio of disc luminosity to corona luminosity
	\item[e] The running time of simulations
\end{tablenotes}
\end{threeparttable}
\end{table}

\section{Vacuum treatment}\label{Sec:VRS}
We adopted the FVM on an Eulerian grid to construct our numerical model due to its simplicity and maturity in incorporating additional important physical processes. However, FVM typically assumes the absence of vacuum in the fluid, and a density floor was introduced to mimic the vacuum, preventing computational crashes when vacuum regions inevitably appear. In the disc-corona interactions, the cold, dense disc gas and hot, tenuous coronal gas coexist, with a density range spanning several orders of magnitude. Given the finite precision of the machine, the resolution of density in simulations is also limited. Approaching the density floor can result in spurious calculations of pressure and sound speed (critical components in FVM), which may lead to erroneous numerical results or even computing crashes due to machine truncation errors. Moreover, gravitational effects inevitably lead to vacuum formation in some CCs containing tenuous gas. If the low-density mimic vacuum is the only option available, this may cause physically inconsistent gas heating and lead to the continuous, spurious generation of mass from cells with the mimic vacuum, severely distorting or even destroying the simulation results. To address this issue, \cite{Munz1994} introduced the GVI tracking algorithm based on the vacuum Riemann problem (VRP). This algorithm resolves the numerical difficulties associated with vacuum regions. Further improvements were made by \cite{Subramaniam2018} for 2D MHD simulations. We adopted their GVI tracking algorithm in Athena++ as an essential numerical technique for our simulations. Readers focused on the physical aspects can skip this section, as it does not affect the interpretation of our numerical results.

\subsection{Vacuum Riemann solver}
The Riemann solver (RS) is a crucial component of many FVM codes \citep{Toro2009}. General RSs are constructed based on the Riemann problem (RP) for two adjacent non-vacuum fluid states. The most widely implemented solver is the HLLC \citep{Toro2009}, which is also available in Athena++ for both HD and MHD. When one of the two fluid states becomes vacuum, the wave propagation around the GVI must be re-analysed using the VRP, which differs from the general RP. Without loss of generality, we considered the VRP with an initial non-vacuum left state and a vacuum right state. \cite{Munz1994} analysed this VRP and presented the exact solution \citep[see also][]{Toro2009}, which can be expressed as
\begin{gather}
	\mathbf{U}(x,t)=\left\{
	\begin{aligned}
		&\left(\rho_{\rm{L}},\ m_{\rm{L}},\ e_{\rm{L}}\right) &\rm{for} &\ x/t<S_{\rm{L}}, \label{eq_cons1}\\
		&\left(\rho_0,\ m_0,\ e_0\right) &\rm{for} &\ S_{\rm{L}}\leq x/t \leq S_{\rm{R}}, \\
		&(0,\ 0,\ 0) &\rm{for} &\ x/t>S_{\rm{R}},
	\end{aligned}
	\right.
\end{gather}
with
\begin{subequations}
\begin{gather}
	S_{\rm{L}}=v_{\rm{L}}-c_{\rm{L}}, \\
	S_{\rm{R}}=v_{\rm{L}}+\frac{2c_{\rm{L}}}{\gamma-1}, \\
	m_{\rm{L}}=\rho_{\rm{L}} v_{\rm{L}},\ e_{\rm{L}}=p_{\rm{L}}/(\gamma-1)+\frac{1}{2}\rho_{\rm{L}} v^2_{\rm{L}}, \label{eq_cons2}\\
	m_0=\rho_0 v_0,\ e_0=p_0/(\gamma-1)+\frac{1}{2}\rho_0 v^2_0, \\
	v_0=\left[(\gamma-1)v_{\rm{L}}+2(x/t+c_{\rm{L}})\right]/(\gamma+1), \\
	\rho_0=\left[\left(v_0-x/t\right)^2 \rho^{\gamma}_{\rm{L}}/(\gamma p_{\rm{L}})\right]^{1/(\gamma-1)}, \\
	p_0=\left(\rho_0/\rho_{\rm{L}}\right)^{\gamma}p_{\rm{L}}.
\end{gather}
\end{subequations}
Here, $\mathbf{U}(x,t)$ represents the vector of conserved quantities, with the subscript ‘L’ denoting the initial non-vacuum left state. Based on this solution, \cite{Munz1994} also developed the VRS, which was later reformulated by \cite{Subramaniam2018} to compute the numerical flux at the GVI for use in FVM codes. The formula for this VRS is given by
\begin{gather}
	\mathbf{F}_{\rm{gv}}=\left\{
	  \begin{aligned} \label{eq_VRS}
	  	&\mathbf{F}_{\rm{L}} &\ \rm{if}\ & S_{\rm{L}}\geq 0, \\
	  	&\frac{S_{\rm{R}} \mathbf{F}_{\rm{L}}-S_{\rm{R}} S_{\rm{L}} \mathbf{U}_{\rm{L}}}{S_{\rm{R}}-S_{\rm{L}}} &\ \rm{if}\ & S_{\rm{L}}<0,
	  \end{aligned}
	\right.
\end{gather}
where $\mathbf{F}_{\rm{gv}}$ is the vector of numerical flux (with the subscript ‘gv’ denoting the GVI), $\mathbf{U}_{\rm{L}}$ is the conserved quantity vector for the initial non-vacuum left state (defined in Eq.~(\ref{eq_cons1})), $\mathbf{F}_{\rm{L}}$ is the flux vector computed using $\mathbf{U}_{\rm{L}}$, and $c_{\rm{L}}$ is the sound speed of the left state. Note that the exact solution of this VRP (Eq.~\ref{eq_cons1}) represents a rarefaction wave, where gas diffuses freely into the vacuum, which is consistent with the physical intuition.

\subsection{Gas-vacuum interface tracking} \label{Sec:GVI}

The implementation of the VRS relies heavily on accurate GVI tracking. The original GVI tracking algorithm by \cite{Munz1994} utilised the exact solution of the VRP to determine the precise position of the GVI within the CCs. This approach effectively confines FVM calculations to non-vacuum regions and avoids computational challenges associated with vacuum. However, this algorithm is limited to 1D FVM calculations, as determining the accurate GVI in complex geometries of 2D or 3D grids is difficult. \cite{Subramaniam2018} relaxed the requirement for precise GVI tracking, assuming that GVIs only appear between CCs, consistent with the basic FVM assumption that discontinuities occur only at CC interfaces. They introduced a density threshold ($\rho_{\rm{t}}$) labelling CCs with densities higher than $\rho_{\rm{t}}$ as gaseous and others as vacuum, even if their densities are slightly below $\rho_{\rm{t}}$. In this scheme, the classical RS (e.g. HLLC) is used to compute flux between two gaseous CCs, while the VRS (Eq.~\ref{eq_VRS}) is applied when the interface lies between a gaseous and a vacuum CC. If both adjacent CCs are labelled as vacuum, flux calculations are skipped, even if they are physically non-vacuum.

\begin{figure}
	\includegraphics[width=\columnwidth]{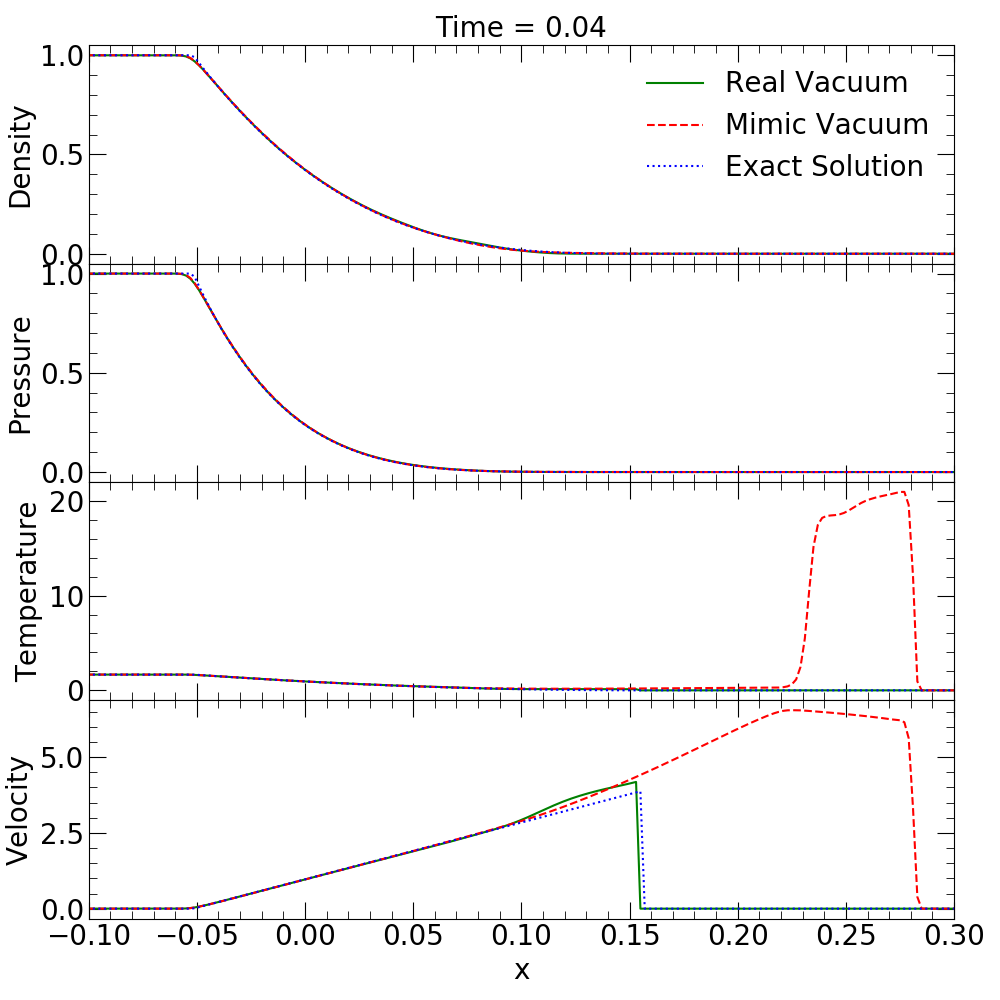}
	\caption{1D simulations of gas freely diffusing into vacuum. For simplicity, all quantities are presented in dimensionless computational units, where both the initial density and pressure are normalised to unity. In this simulation, the left side of $x=0$ is initially a constant non-vacuum state, while the right side is set to a vacuum state. Three solutions are presented: the exact solution, the numerical solution involving real vacuum, and the numerical solution involving mimic vacuum. All curves correspond to the same simulation time. }
	\label{fig1}
\end{figure}

This cell-labelling algorithm from \cite{Subramaniam2018} depends solely on the properties of the CCs and is independent of the grid geometry, making it compatible with any Eulerian grid. Consequently, it can be easily integrated into any FVM code. We have incorporated this algorithm and VRS (Eq.~\ref{eq_VRS}) into the HD part of Athena++'s HLLC solver (the MHD implementation will be made in our future work). This extension allows FVM calculations involving a vacuum from 1D to 3D. To test and validate the enhanced solver, we ran simulations of a 1D gas diffusion into a vacuum with both real-vacuum and mimic-vacuum right states. The results, in Fig.~\ref{fig1}, show that both numerical methods are consistent with the exact solution for density and pressure. However, the mimic-vacuum case exhibits significant deviations in temperature and velocity, while the GVI tracking and VRS method closely matches the exact solution. In the relevant simulations, the density threshold for the GVI tracking algorithm was set to the same low-density value as in the mimic-vacuum case (both were set to $10^{-10}$ times the initial left-state density), highlighting the impact of different vacuum treatments. The temperature spike in the mimic-vacuum case (see the right part of the temperature panel in Fig.~\ref{fig1}) is due to shock heating from the dense left flow interacting with the tenuous mimic-vacuum gas, a problem first noted by \cite{Subramaniam2018}. Clearly, the improved Athena++ solver using GVI tracking and VRS accurately models the process of gas diffusing into vacuum, resolving the challenges previously associated with a vacuum in FVM calculations.

\section{Numerical results}\label{Sec:Results}
\subsection{Adiabatic model}\label{subSec:Ad_Model}
In our model, the implementation of turbulent viscosity and thermal conduction leads to a significant reduction in the simulation time-step by one to two orders of magnitude, compared to calculations without these processes. To save CPU time, the general approach is to first run an adiabatic calculation (without turbulent viscosity, thermal conduction, and cooling mechanisms) for a period, allowing the system to relax to a quasi-steady state,  in which variations of disc-corona structure are significantly reduced compared to the initial moment \citep[e.g.][]{Nemmen2024}.

All four cases defined in Table~\ref{tab:DiffCases} have been run adiabatically but we only show case A as an example to illustrate this first run. The density threshold for GVI tracking in this simulation is set to $\rho_{\rm{t}} = 10^{-12}\rm{g/cm^3}$ (identical for all models), which is approximately 11 orders of magnitude lower than the maximum initial disc density. The simulation has been run for a physical duration of $\sim 2.45 \times 10^5 r_{\rm{s}}/c$ (we followed \cite{Wu2016} and \cite{Nemmen2024} in using the $r_{\rm{s}}/c$ as time unit, which facilitates the comparison of our numerical results with other research works), consuming about two days of CPU time. This corresponds to $\sim 78.24$ ($\sim 874.73$) periods of Keplerian rotation at the outer boundary, $r = 50r_{\rm{s}}$ (inner boundary $r = 10r_{\rm{s}}$), which is a sufficiently long relaxation time to reach quasi-steady conditions. Although the first run is adiabatic, the higher temperature of the coronal gas compared to the disc gas has been observed in the quasi-steady state (see the right half panels in Fig.~\ref{fig2} and low half ones in Fig.~\ref{fig3}). Therefore, we need to verify that the observed temperature increase is physically reasonable rather than resulting from spurious numerical heating. 

\subsubsection{Acoustic shock heating}\label{subsubsec:Acoustic_sho_hea}
\begin{figure}
	\includegraphics[width=\columnwidth,trim=440 1835 350 0, clip]{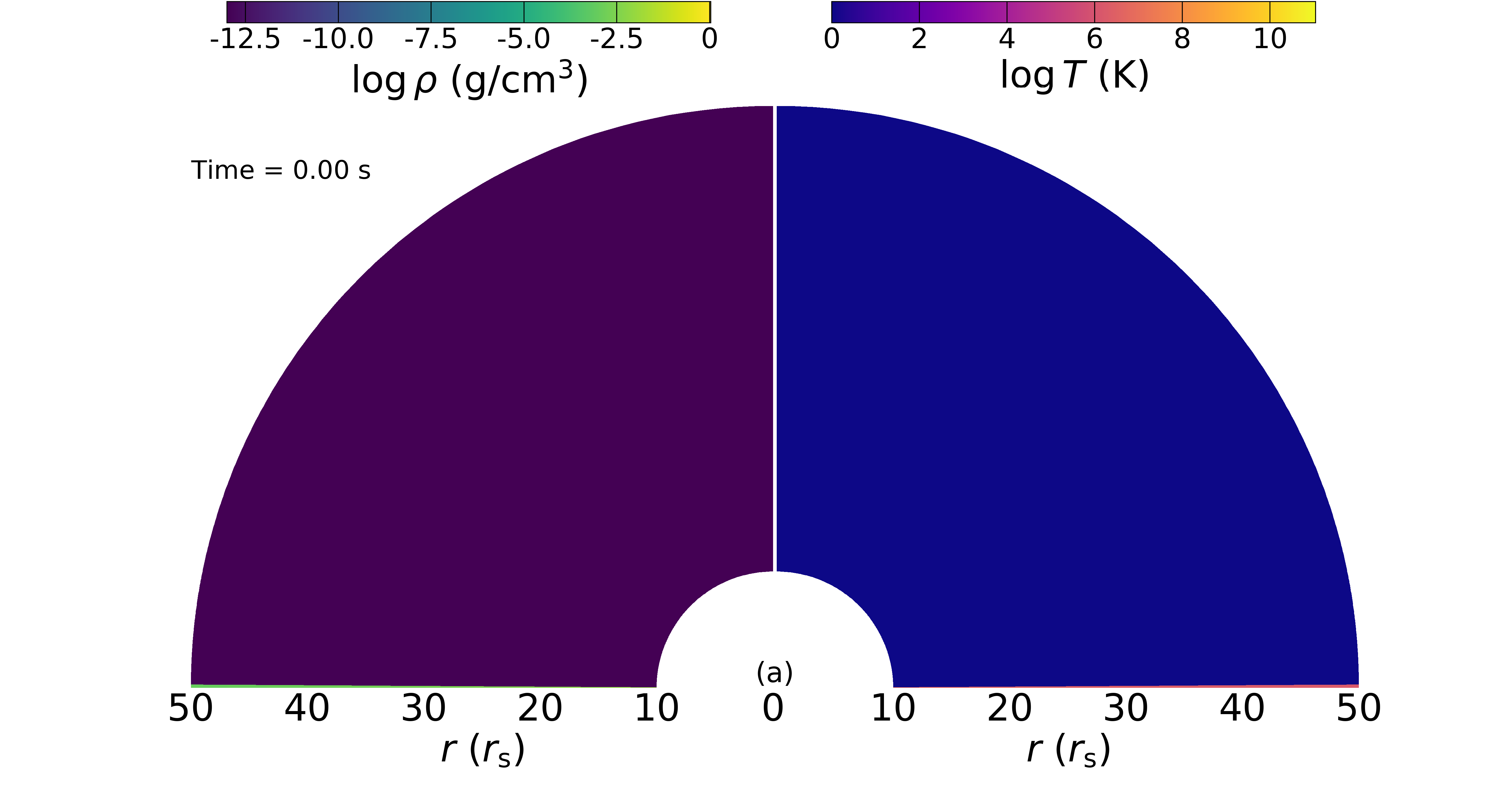}\\
	\includegraphics[width=\columnwidth,trim=440 262 350 1700, clip]{fig2a.png}\\
	\includegraphics[width=\columnwidth,trim=440 215 350 1550, clip]{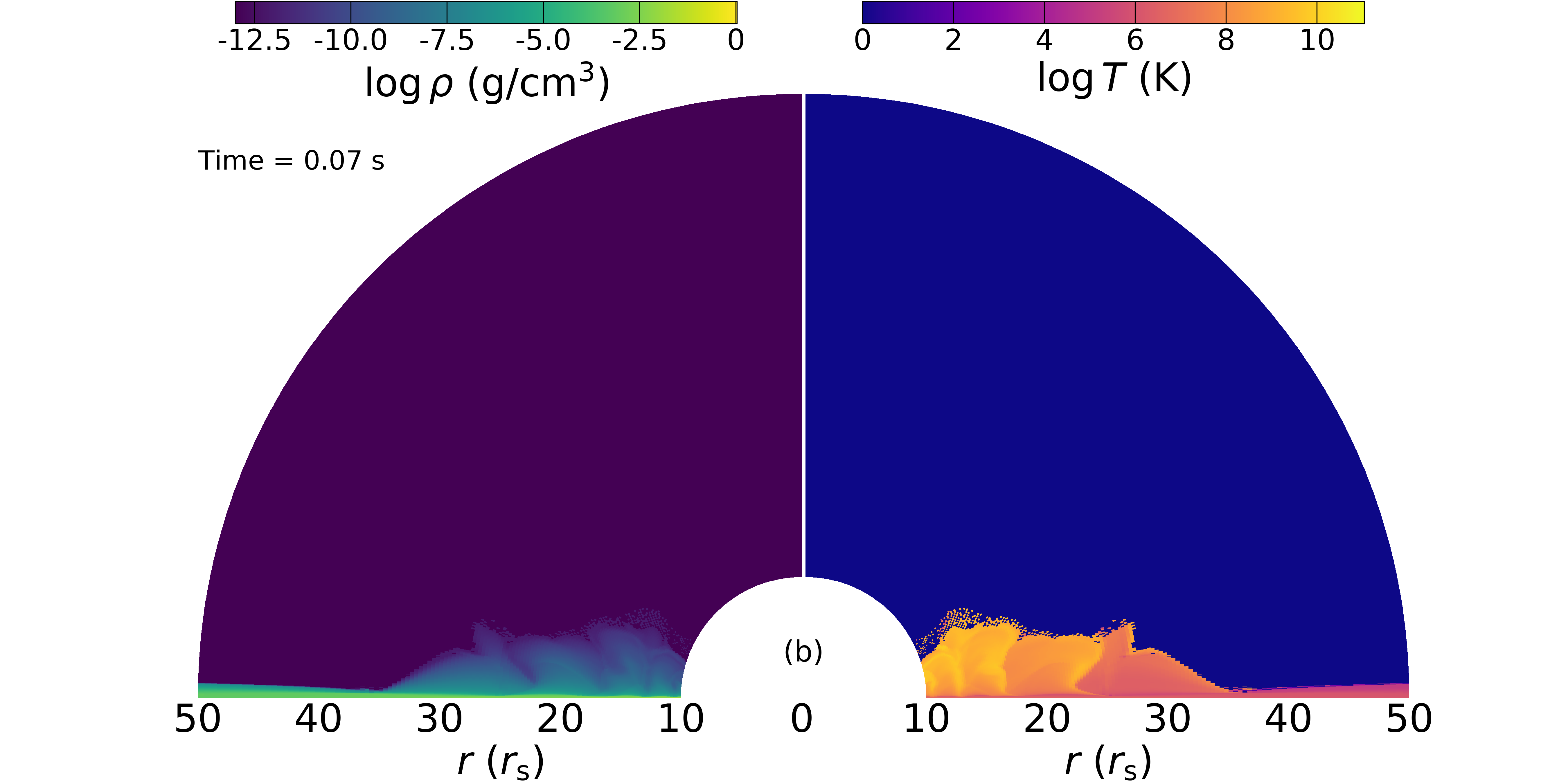}\\
	\includegraphics[width=\columnwidth,trim=440 215 350 1300, clip]{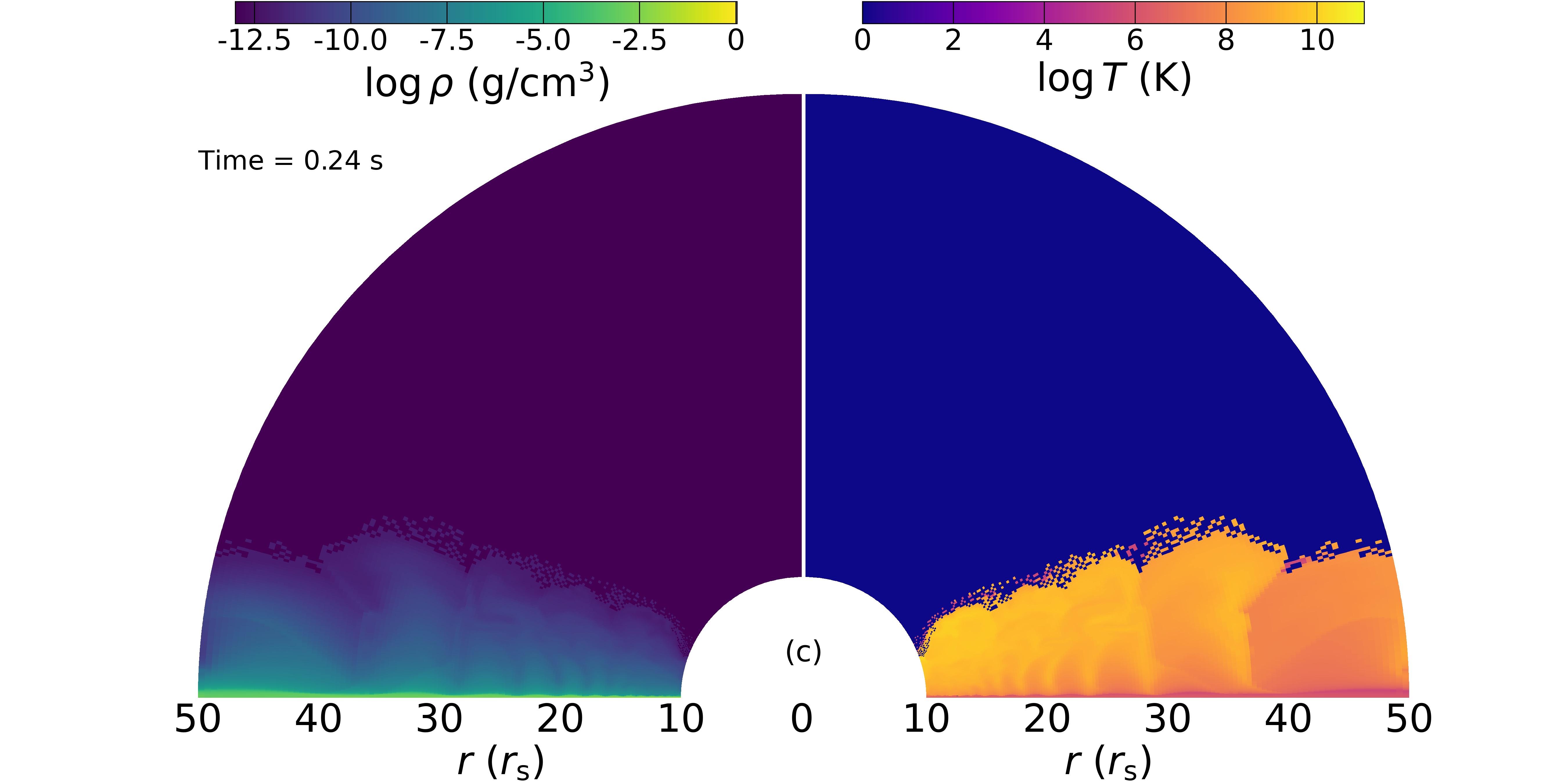}\\
	\includegraphics[width=\columnwidth,trim=440 10 350 1150, clip]{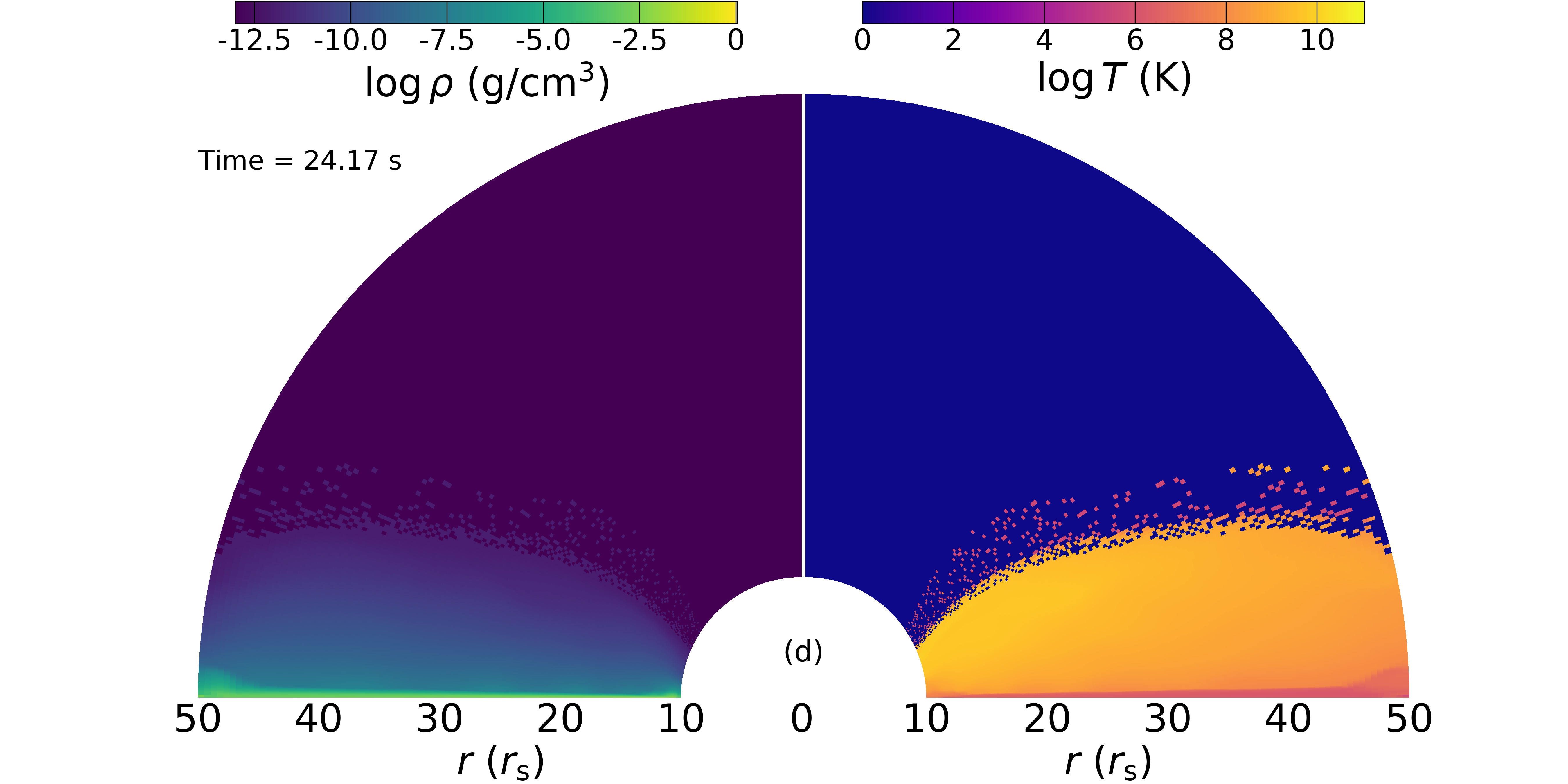}
	\caption{Pseudo-colour maps of the 2D distributions of density and temperature for case A at four special moments: $0$ (a), $7.11 \times 10^2$ $r_{\rm{s}}/c$ (b), $2.44 \times 10^3$ $r_{\rm{s}}/c$ (c), and $2.45 \times 10^5$ $r_{\rm{s}}/c$ (d). In panel (d), the model has reached the quasi-steady state. Corresponding equatorial close-up views for these panels are shown in Fig.~\ref{fig3}.}
	\label{fig2}
\end{figure}

\begin{figure}
   \includegraphics[width=\columnwidth]{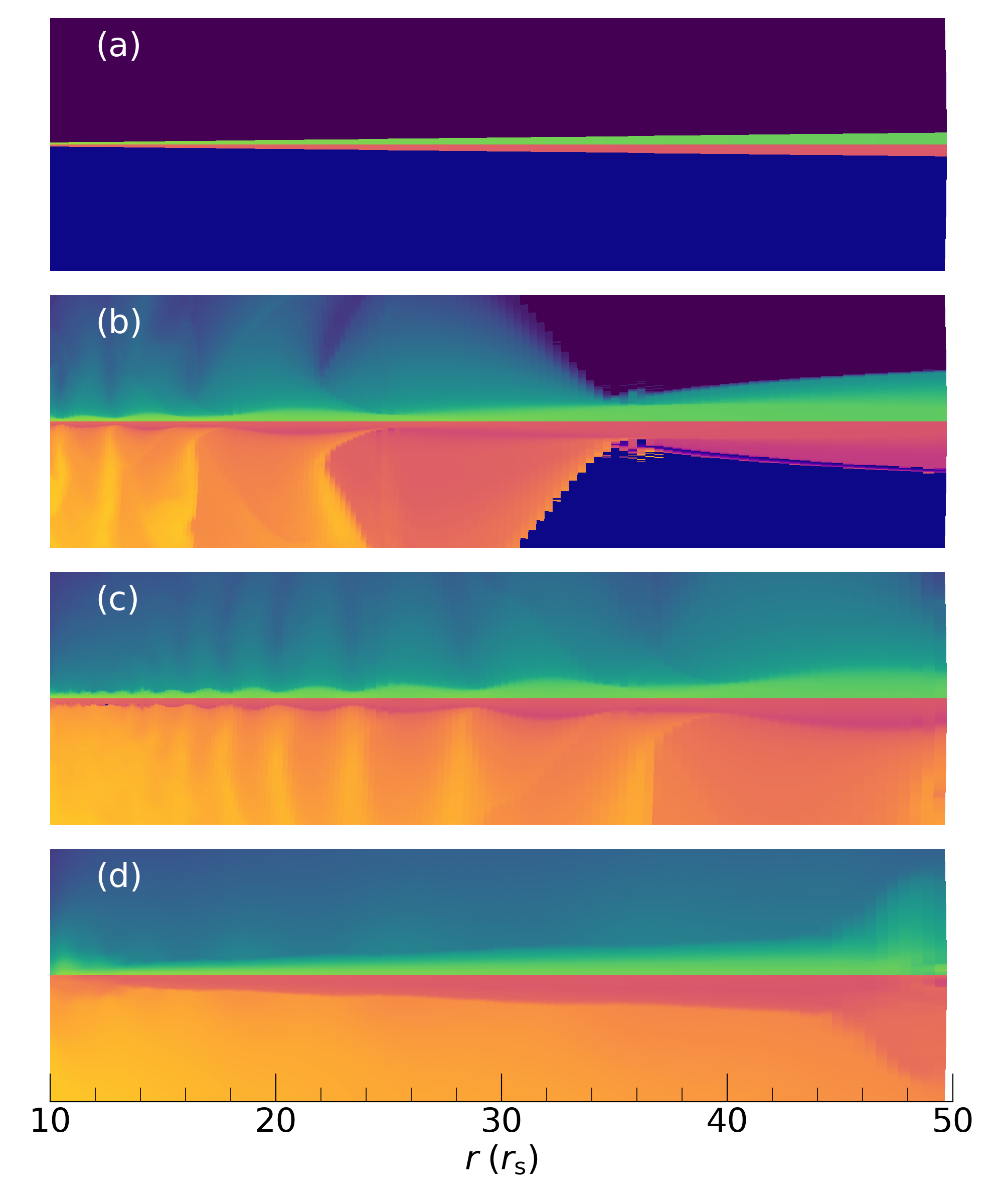}
   \caption{Equatorial close-up views of the density (upper of each panel) and temperature (lower of each panel) distributions near the equatorial plane in Fig.~\ref{fig2}. The colours are scaled using the same colour bar as in Fig.~\ref{fig2}.}
   \label{fig3}	
\end{figure}

In Fig.~\ref{fig2}, we present pseudo-colour maps of the density and temperature distributions at four selected moments for case A\footnote{We consider only gas pressure and ignore radiation pressure when calculating the temperature for plotting. This may overestimate the temperature in optically thick regions but has no effect on our simulations, as this is merely a post-processing step.} (see also Fig.~\ref{fig3} for the corresponding equatorial close-up views). Panel (a) represents the initial state consisting of a thin disc and vacuum, while panel (d) corresponds to the final state, where some hot gas, with temperatures reaching up to $\sim10^{10}$ K, rises from the disc and reaches a quasi-steady state above the disc. 

In the right halves of panels (b) and (c) in Fig.~\ref{fig2} (see also the lower halves of panels (b) and (c) in Fig.~\ref{fig3}), it is evident that a wave train propagates outwards, accompanied by significant temperature increases at each shock contact surface, where the mechanical energy carried by the waves is dissipated through shock interactions, thereby heating the gas. These waves may be triggered by differential rotation in the accretion disc near the inner boundary and continue to heat the coronal gas via shock dissipation during the quasi-steady state. It is important to note that the adiabatic expansion of disc gas into a vacuum cannot directly generate hot gas in our model, as is shown by the thickening of the undisturbed disc in panels (b) and (c) of Fig.~\ref{fig2} (see also panels (b) and (c) in Fig.~\ref{fig3}). This result is ensured by the vacuum algorithm. Therefore, the only plausible source of heating is shock dissipation caused by the colliding waves. Given the adiabatic nature of our model, gas outflow through the inner and outer unidirectional boundaries plays a role in cooling, achieving equilibrium with the shock heating process. We believe that this heating mechanism is acoustic shock heating, which is widely recognised as an effective chromospheric HD heating process in the solar physics community \citep{Narain1996, Aschwanden2005, Erdelyi2007}. 

\subsubsection{Quasi-steady state of disc and corona}\label{subsubsec:quasi}
\begin{figure}
	\includegraphics[width=\columnwidth]{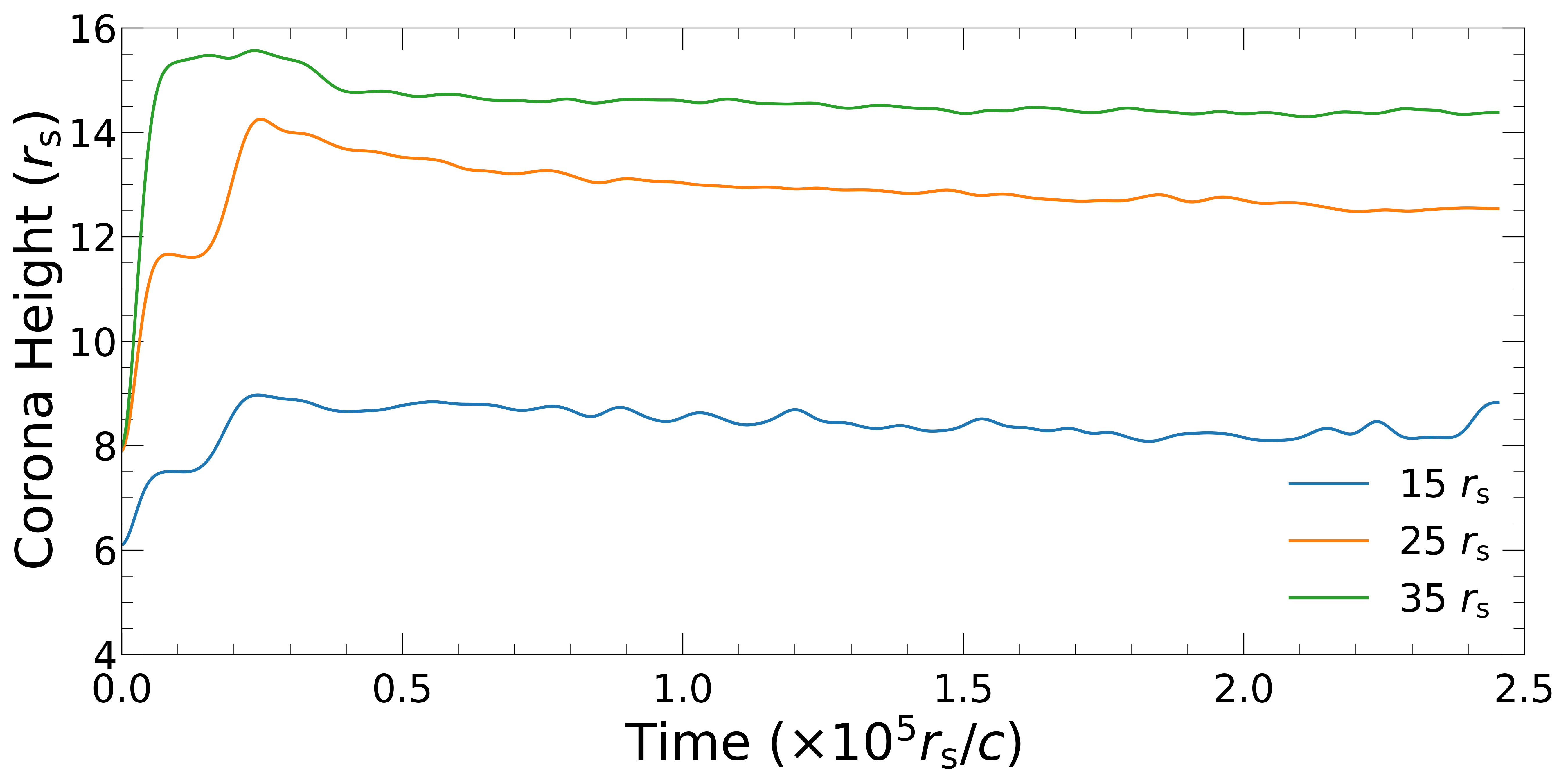}
	\caption{Time evolution of the corona height at different radii ($15\ r_{\rm{s}}$, $25\ r_{\rm{s}}$, and $35\ r_{\rm{s}}$) for case A during the adiabatic relaxation process.}
	\label{fig4}
\end{figure}

At $\sim 2.45 \times 10^5 r_{\rm{s}}/c$, all four models have already reached a quasi-steady state for a period of time. Since the density and temperature distributions are similar across the models at this moment, only case A is presented for simplicity; it is shown in panel (d) of Fig.~\ref{fig2}. To confirm that the model has indeed reached a quasi-steady state, Fig.~\ref{fig4} shows the time evolution of the corona height at $15r_{\rm{s}}$, $25r_{\rm{s}}$, and $35r_{\rm{s}}$ for case A (for details on how we distinguish the disc, corona, and vacuum, see Sect.~\ref{subsubsec:strctrue}). Initially, the corona height at all radii increases rapidly, followed by a gradual stabilisation. This behaviour suggests that the corona geometry evolves dynamically in the early stages, but reaches a quasi-steady state after $\sim 5\times 10^4 r_{\rm{s}}/c$, with the final corona height positively correlated with the radius.

\subsection{Disc-corona interaction}
Using the quasi-steady results from the adiabatic simulations as the initial condition (see panel (d) in Fig.~\ref{fig2}), we initiated full simulations by incorporating turbulent viscosity, thermal conduction, bremsstrahlung cooling, and artificial disc cooling (all described in Sect.~\ref{Sec:Model}) to investigate the disc–corona interaction. Here, we present four typical cases and discuss their results. The parameters and some relevant results are listed in Table~\ref{tab:DiffCases}. These cases share the same radii of the inner and outer boundaries ($r_{\rm{in}} = 10r_{\rm{s}}$ and $r_{\rm{out}} = 50r_{\rm{s}}$). The differences lie in the viscosity parameter ($\alpha$) and the initial accretion rate ($\dot{m}$), which are varied to explore their effects on disc evaporation. In addition, the computational time-step is strongly influenced by viscosity so that the time-steps of these four cases are too small to allow simulations to reach the inflow-outflow equilibrium within a tolerable time. Therefore, the running time of these four cases is different (see the last line in Table~\ref{tab:DiffCases}), and we cannot determine which case would result in complete disc evaporation, but we can illustrate the dynamic process of evaporation. We notice that the simulations of \cite{Wu2016} and \cite{Nemmen2024} also suffer from similar situations.

\begin{figure}
	\includegraphics[width=\columnwidth]{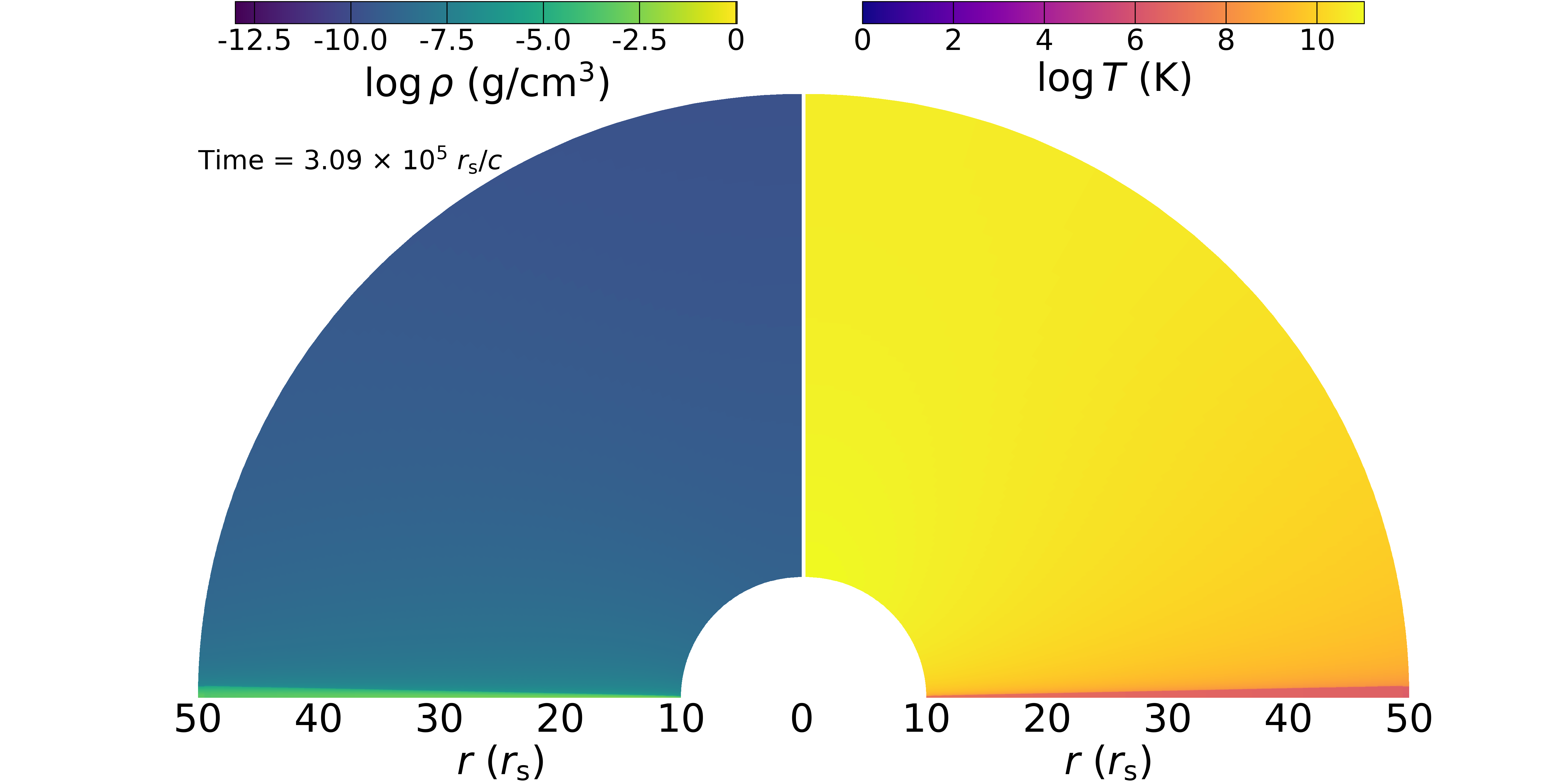}
	\caption{Temperature and density distributions for case A at the last moment ($\sim 3.09 \times 10^5 r_{\rm{s}}/c$) in the simulation. The corona gas has filled the entire computational domain above the disc flow.}
	\label{fig5}
\end{figure}

In Fig.~\ref{fig5}, we present the temperature and density distributions for case A at the last moment of the simulation. It can be observed that the hot and tenuous coronal gas fills most of the computational domain, while a cold and dense disc structure remains near the equatorial plane. The temperature and density distributions are visually similar across the four cases; therefore, further analysis is required and is provided in the subsequent part of this subsection.

\subsubsection{Disc evaporation}\label{subsubsec:DiscEvap}
\begin{figure}
	\includegraphics[width=\columnwidth]{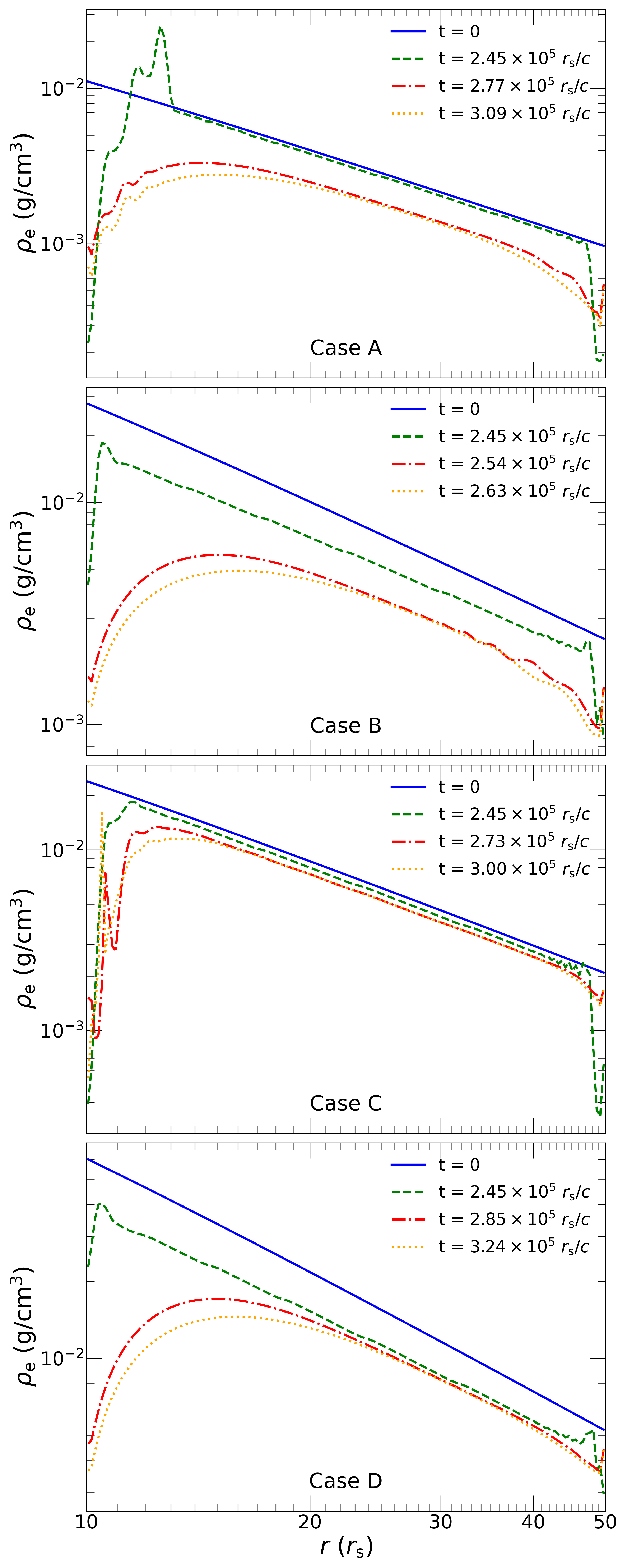}
	\caption{Profiles of the equatorial density at different moments for all cases. The solid blue line represents the time when the relaxation process ends, and the dash-dotted orange line corresponds to the last moment for the respective case.}
	\label{fig6}
\end{figure}
To reveal the different degrees of disc evaporation, we used the equatorial density ($\rho_{\rm{e}}$) as the evaporation indicator, which is generally the densest in the local disc. In Fig.~\ref{fig6}, we show the profiles of $\rho_{\rm{e}}$ at different moments for four typical cases, in which the extent of the density reduction can reflect the level of disc evaporation. The main focus in these profiles is on the $\rho_{\rm{e}}$ in the middle radial region, because it is away from the boundaries and less affected by the artificial BCs, presenting more authentic evaporation behaviour. The high-viscosity cases, A and B, reveal more intense evaporation than the low-viscosity cases, C and D (the maximum $\rho_{\rm{e}}$ on the dotted orange lines of cases A and B are approximately one order of magnitude lower than the ones of cases C and D), which presents a significant effect brought by viscous heating and is consistent with the DEM \citep[see][]{Liu2022}. After adding viscous heating, the evaporation rates of all four cases are initially fast and then slow down (see the obviously different distances between the dashed green, dash-dotted red, and dotted orange lines over equal time spans), and the evaporation is faster at inner radii than outer ones (see the significant drop in the dash-dotted red and dotted orange lines within $20 r_{\rm{s}}$). At the same time, the influence of the accretion rate can also be seen from the comparison between cases A and B, and between cases C and D. These comparisons both show that the evaporation is stronger at high accretion rates than at lower ones (see the separation between the dotted orange line and the solid blue line of each case, which reflects the extent of disc evaporation), indicating that the excessive accumulation of disc material caused by high accretion rates is almost evaporated under the same viscous heating conditions, and the amount left depends more on the viscosity (more viscosity leads to less disc material left) rather than the accretion rate.

\begin{figure}
	\includegraphics[width=\columnwidth]{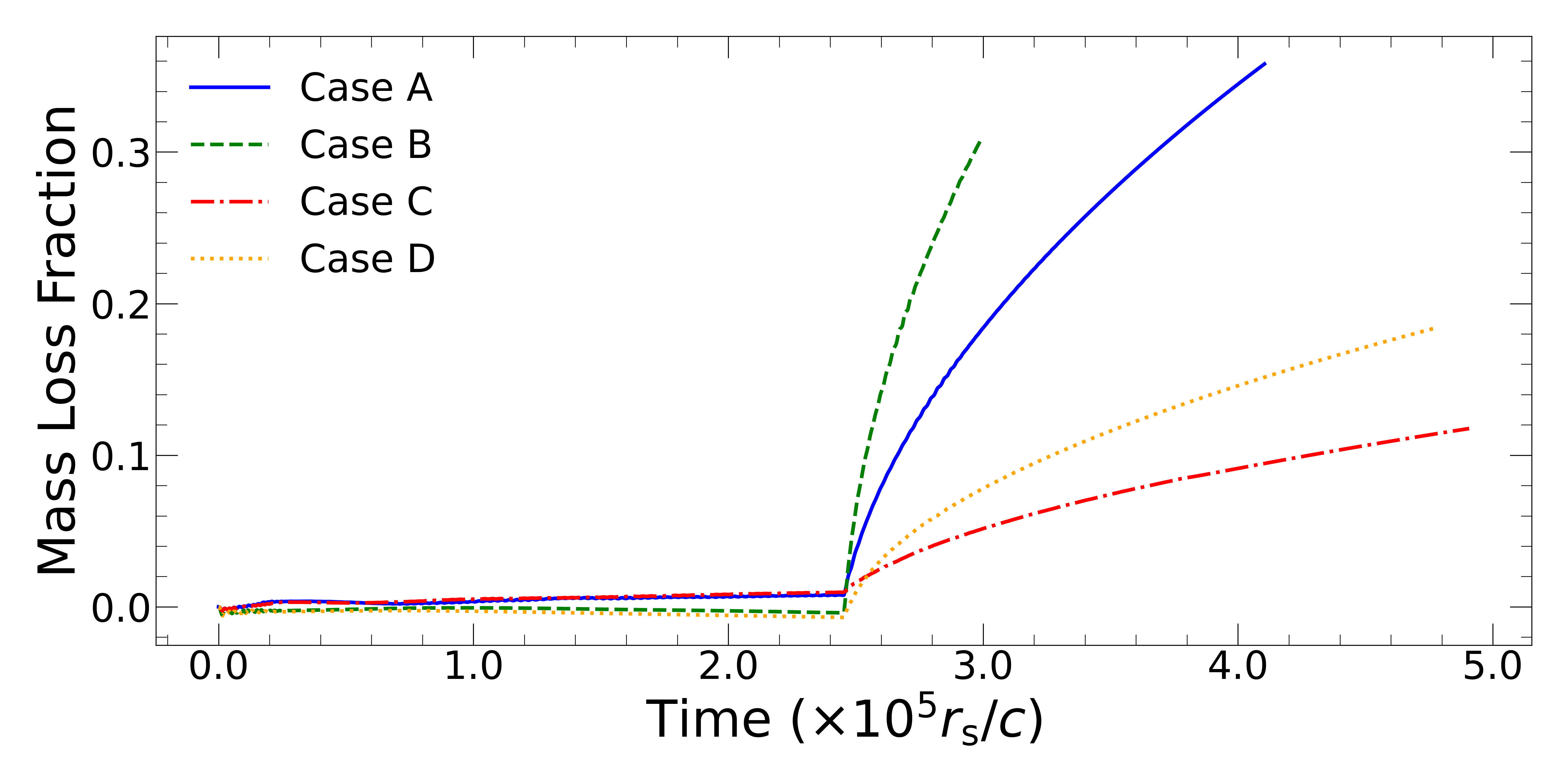}
	\caption{Mass loss fraction over time for all cases, compared to the initial disc mass at $t=0$. The first $2.45 \times 10^5 r_{\rm{s}}/c$ is in the adiabatic phase, after which cooling and heating mechanisms are introduced. The different lengths of these lines correspond to varying running times (see the last line in Table~\ref{tab:DiffCases}).}
	\label{fig7}
\end{figure}

In Fig.~\ref{fig7}, we present the mass loss fraction (which is defined as the ratio of total cumulative mass loss to the initial disc mass) curves for four cases. The value of this fraction can reflect the specific intensity of the evaporation rate. During the adiabatic phase ($t<2.45 \times 10^5 r_{\rm{s}}/c$), the evaporation rates of cases A (solid blue line) and C (dash-dotted red line) with $\dot{m}=0.01$ are very close and both higher than the ones of cases B (dashed green line) and D (dotted orange line) with $\dot{m}=0.1$; the rates of cases B and D are also very close to each other. After $t=2.45 \times 10^5 r_{\rm{s}}/c$, these curves are clearly separated from each other and significantly increase to a higher level than the ones in the adiabatic phase due to the addition of viscous heating. This reveals that a higher viscosity or accretion rate results in a higher evaporation rate, with the viscosity taking priority over the accretion rate (e.g. the evaporation rate of case A is higher than that of case D, despite its lower accretion rate, highlighting the dominant role of viscosity.). This is consistent with the disc evaporation characteristics presented in Fig.~\ref{fig6}. 

It is worth mentioning that enabling both thermal conduction and turbulent viscosity in our simulations increases the numerical time-step by $\sim 3$ orders of magnitude compared to viscosity-only cases. Thermal conduction dissipates internal energy, preventing localised temperature spikes that would otherwise drastically reduce the viscous time-step. Thus, even with a much higher viscosity ($\alpha=0.3, 0.9$) than \cite{Wu2016} and \cite{Nemmen2024} ($\alpha=0.01$), our simulations can still achieve similar peak temperatures ($\sim 10^{11}$ K) and barely manageable time-steps ($\sim 10^{-4}r_{\rm{s}}/c$, but it would be $\sim 10^{-7}r_{\rm{s}}/c$ in viscosity-only cases). Due to these extremely small time-steps, our full simulations are too slow to reach inflow-outflow equilibrium in a tolerable time. However, this reflects a non-equilibrium of disc evaporation, which is different from the disc-condensed ones of \cite{Wu2016} and \cite{Nemmen2024}.

\subsubsection{Disc-corona structure}\label{subsubsec:strctrue}
Since there are vacuum CCs in our algorithm, we use the density threshold, $\rho_{\rm{t}}$ (see Sect.~\ref{subSec:Ad_Model}), to outline the surface of disc-corona structure in the vacuum. Computational cells with a density below this threshold are classified as part of the vacuum region, while the remaining cells are considered part of the disc-corona structure. Similarly, the interface between disc and corona can also be outlined according to the actual cooling rate implemented in CCs (see Sect.~\ref{subsubsec:rad_cooling}). If the bremsstrahlung radiation cooling is implemented in a CC, this cell will be classified as part of the corona. Otherwise, it will be regarded as a cell of disc flow.

\begin{figure}
	\includegraphics[width=\columnwidth]{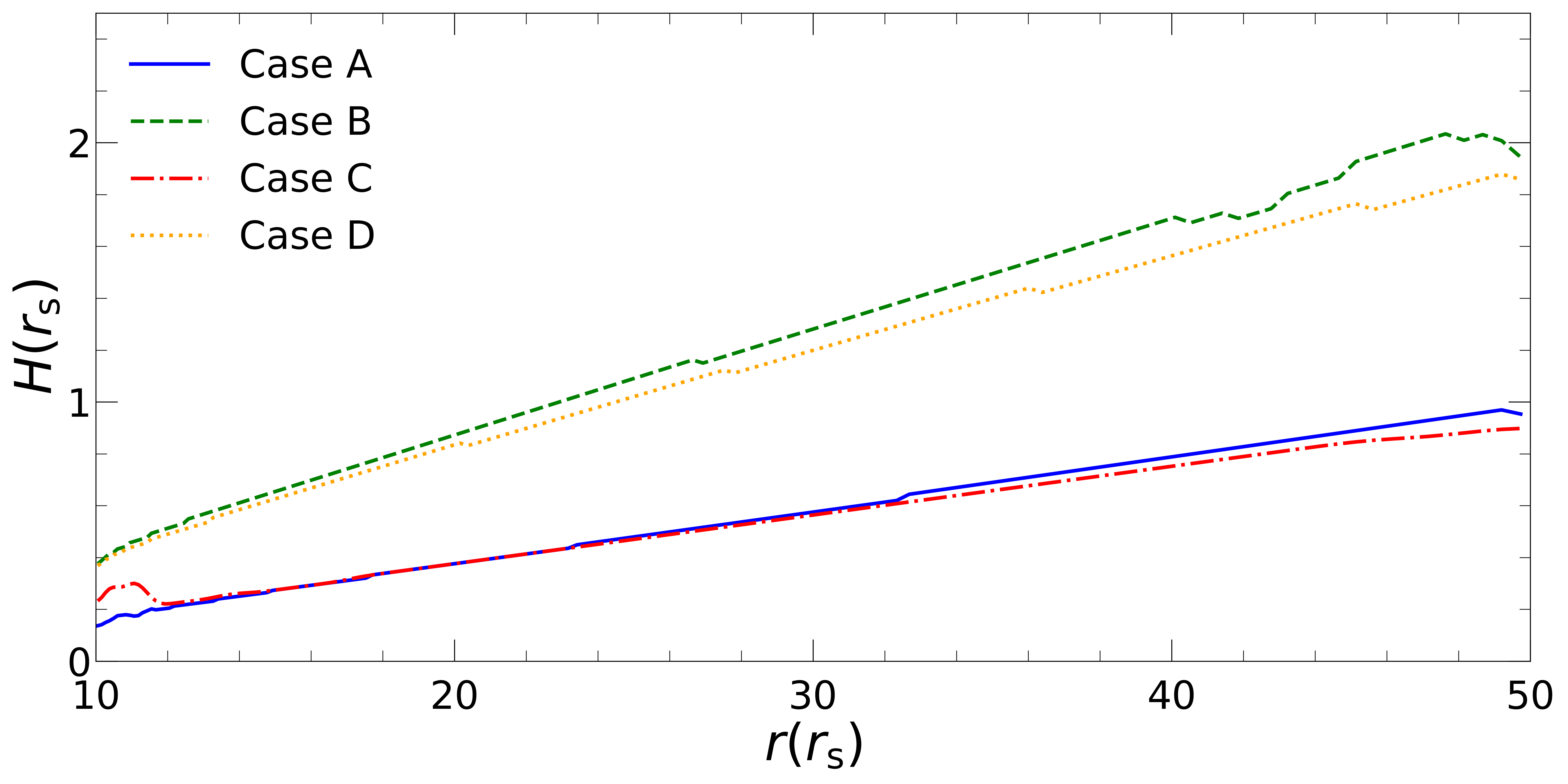}
	\caption{Profiles of disc half-thickness for four cases at the last moment of simulations.}
	\label{fig8}
\end{figure}

In Fig.~\ref{fig8}, we show the profile of disc half-thickness determined by the above-mentioned method at the last moment of simulations for the four cases. This figure reveals that the disc is thinner near the black hole and becomes thicker at larger radii. This behaviour is primarily driven by the distribution of thermal pressure and angular momentum transport within the disc. For cases B and D with higher accretion rates, the disc half-thicknesses are larger than the ones in cases A and C. The increased accretion rate leads to a higher energy deposition in the outer disc regions, which raises the thermal pressure and causes the disc to puff up. This positive correlation between the disc half-thickness and the accretion rate is consistent with the description of the SSD model (described in Chapter 3.2 of \citet{Kato2008}).

\begin{figure}
	\includegraphics[width=\columnwidth]{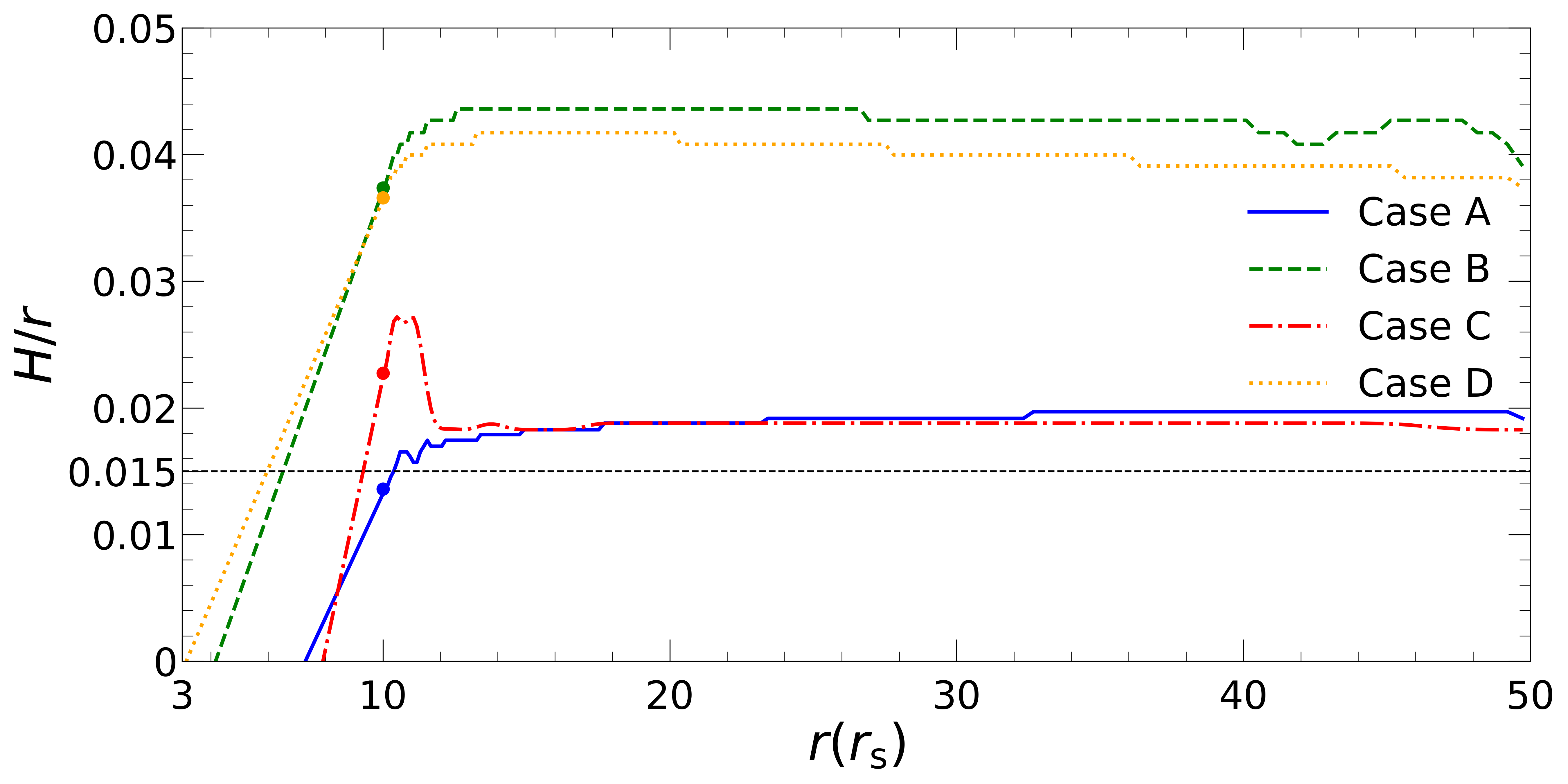}
	\caption{Profiles of the disc half-thickness to radius ratio ($H/r$) at the last moment for all cases. The curve from 10$r_{\rm{s}}$ to 50$r_{\rm{s}}$ derives from our simulation, while the curve from 3$r_{\rm{s}}$ to 10$r_{\rm{s}}$ comes from the linear extrapolation. The starting points of extrapolated lines on the $H/r$ curves are marked with solid coloured circles. The dashed black line represents  $H/r = 0.015$, which is the criterion for determining the truncation radius.}
	\label{fig9}
\end{figure}

We also plot the curve of \( H/r \) as a function of radius in Fig.~\ref{fig9}. Given that our inner boundary is at 10$r_{\rm{s}}$, we performed a linear extrapolation based on the region where \( H/r \) changes significantly (from 10$r_{\rm{s}}$ to 10.5$r_{\rm{s}}$) and extended the curve to the vicinity of the innermost stable orbit (about 3$r_{\rm{s}}$). Using the truncation radius criterion from \cite{Nemmen2024}, where $H/r \leq 0.015$ (the horizontal dashed black line in Fig.~\ref{fig9}), we obtained the truncation radii for each case, shown by the $r_{\rm{tr}}$ values in Table~\ref{tab:DiffCases}. This extrapolation approach represents a last resort to estimate the truncation radii. Therefore, the obtained values can only be regarded as the upper limit estimates of the actual truncation radii under the influence of inner BCs. However, Fig.~\ref{fig9} shows that the accretion rate dominates the truncation of the accretion disc rather than the viscosity (cases A and C, which have low accretion rates, have similar truncation radii, while cases B and D with high accretion rates also show comparable truncation radii) , and the lower accretion rate leads to the larger truncation radius, which is qualitatively consistent with the results of DEM \citep{Liu2022} as well as simulations of \cite{Nemmen2024}. It is worth mentioning that since our model is far from the mass inflow-outflow equilibrium, the truncation radius presented here reflects an instantaneous state during disc evaporation, which is different from the steady truncation radius predicted by DEM.

\subsubsection{Warm corona}\label{subsubsec:warm_coro}
The dichotomy between the disc and corona mentioned above is overly artificial and simplistic. The actual disc-corona structure is likely far more complex. For example, in many Type 1 AGNs, extrapolation of the $2$–$10$ keV power-law down to the soft X-ray band below $2$ keV often reveals an excess of emission. To explain this excess, the warm corona model has been proposed as a possible observational interpretation, requiring the presence of warm gas with the temperature of $0.1$–$2$ keV and the optical depth of $10$–$20$ \citep{Petrucci2020}. This kind of excess has also been observed in BHXBs. For instance, \cite{Jin2024} suggested that the soft X-ray excess during a super-Eddington outburst of 4U 1543-47 in 2021 likely originated from Comptonisation by a warm corona with a temperature of $\sim2$ keV.

\begin{figure}
	\includegraphics[width=\columnwidth]{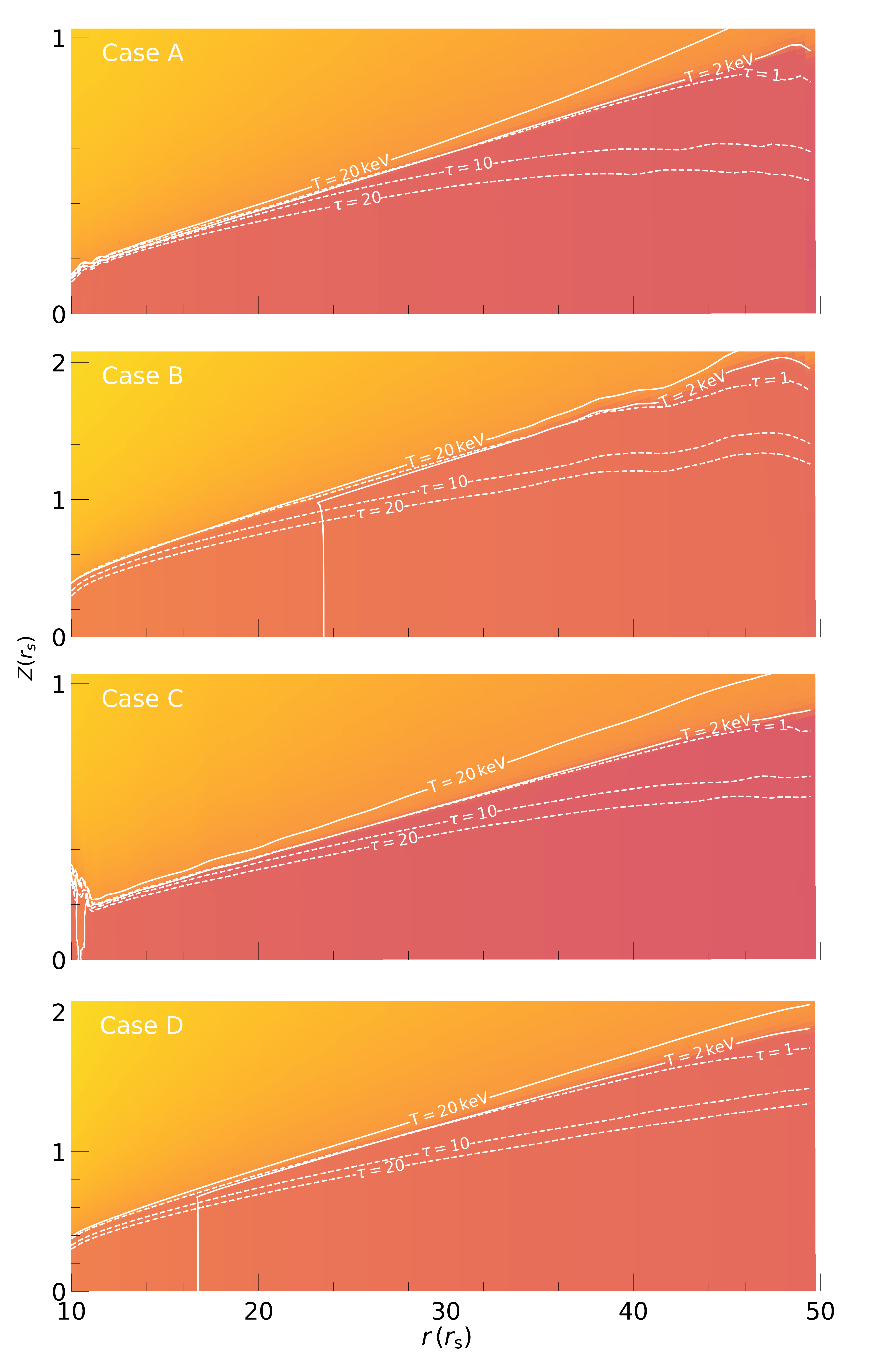}
	\caption{Various characteristic contours overlaid on the background of temperature distribution map (colour-coded with the same colour bar as that of Fig.~\ref{fig2}) near the equatorial plane for the last moments of four cases. The characteristic contours are for the temperatures $T=2$, $20$ keV (solid lines) and the vertical optical depth $\tau=1$, $10$, $20$ (dashed lines), respectively. }
	\label{fig10}	
\end{figure}

Since our simulations involve the interaction between a cold disc and a hot corona, they may contain the warm gas required by the warm corona model. Fig.~\ref{fig10} displays various characteristic contours overlaid on the background of the temperature distribution map near the equatorial plane (the vertical axis of each panel has been stretched). It can be seen that there is always an overlap between the vertical optical depth range of $10$-$20$ (defined as the scattering optical depth for an infinitely distant face-on observer) and the temperature range of $0.1$-$2$ keV for four cases. This indicates that a warm corona indeed exists as a thin layer sandwiched between the hot corona and the cold disc, consistent with the location suggested by \cite{Petrucci2020}. In our four cases, the disc regions ($\tau>20$) uniformly extend inwards to the inner boundary. Therefore, the truncation radii determined in the above subsection are credible and unaffected by the warm corona. Nevertheless, in cases B and D with high $\dot{m}$, the inner region becomes hot ($T>2$ keV), showing a trend toward evaporation truncation. Case C, on the other hand, is somewhat special compared to the others, as it exhibits a truncated cold clump appearing near the inner boundary.

\subsection{Connection to observations}
\subsubsection{Luminous variability}\label{subsubsec:Lum}

\begin{figure}
	\includegraphics[width=\columnwidth]{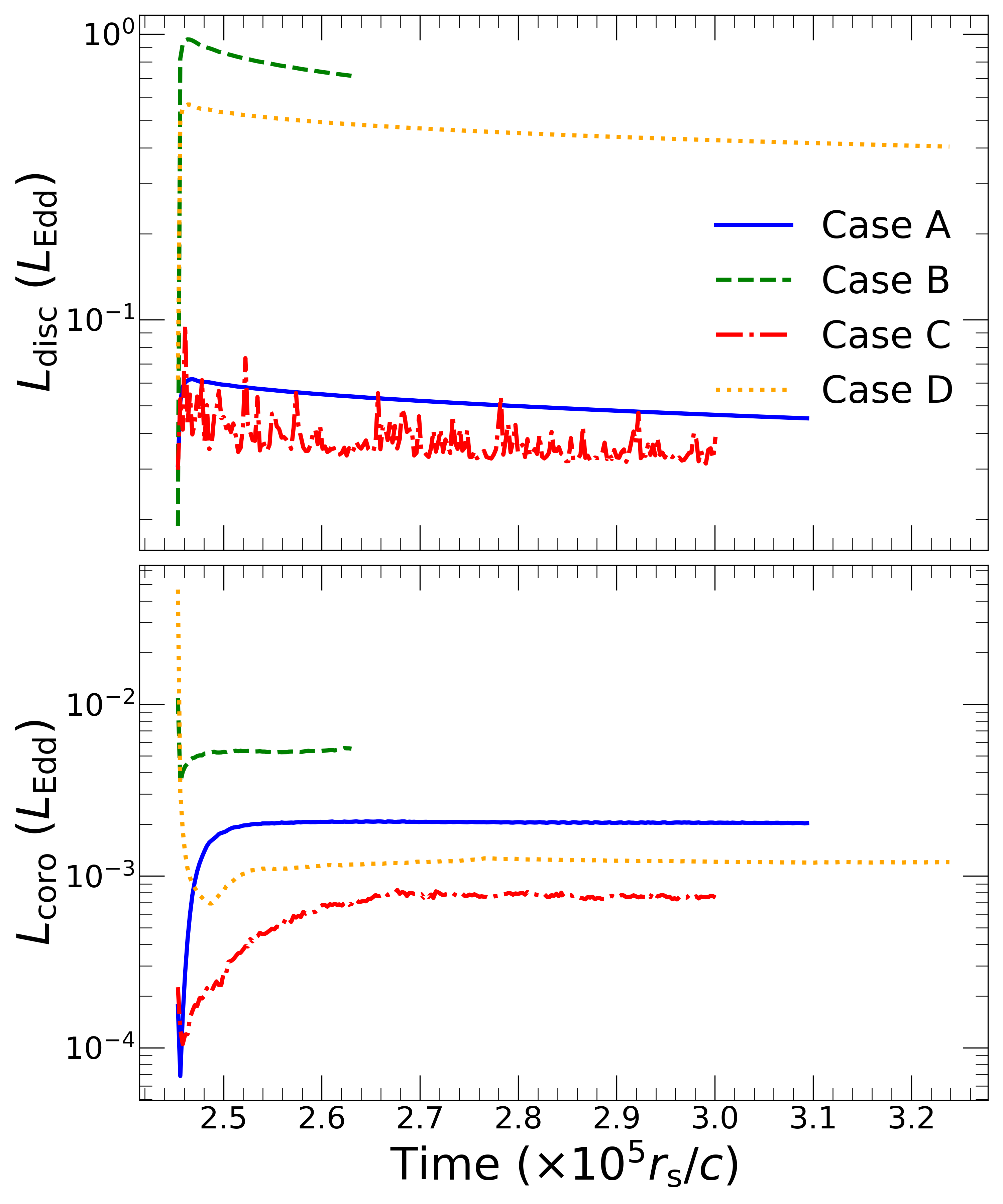}
	\caption{Evolution of disc and corona luminosities for the four cases after the inclusion of the cooling and heating mechanisms.}
	\label{fig11}
\end{figure}

\begin{figure}
    \includegraphics[width=\columnwidth]{fig12.png}
    \caption{Disc truncation radius versus the luminosity. The different point styles represent different observations of GX 339-4 \citep{Miller2006, Reis2008, Kolehmainen2014, Petrucci2014, DeMarco2015, Garcia2015, Plant2015, Basak2016, Wang-Ji2017}, previous simulations \citep{Takahashi2016, Liska2022, Nemmen2024}, and our simulation marked by red asterisks.}
    \label{fig12}
\end{figure}

The observational connection of our numerical results can be presented through the post-processing of numerical data. In this paper, we have generated various mimic light curves, which were achieved by integrating instantaneous cooling rates re-calculated from these data. In Fig.~\ref{fig11}, we show the light curves from the disc and corona for all cases. The disc luminosity was obtained by integrating the artificial cooling rate (Eq.~\ref{eq_artfcoolingrate}), while the corona luminosity was obtained by integrating the bremsstrahlung cooling rate. The distinction between them follows the criterion we mentioned earlier (see Sect.~\ref{subsubsec:strctrue})\footnote{Comparing Fig.~\ref{fig10} with Fig.~\ref{fig8} reveals that the $T=2$ keV isotherm (or $\tau=1$) of each case closely resembles the profile of the disc half-thickness $H$ for the same case (although our method for determining $H$ is not directly based on temperature demarcation). This implies that the warm gas ($T<2$ keV and $1<\tau<20$) is classified as disc gas by our method, suggesting that the obtained $L_{\rm{disc}}$ incorporates contributions from these warm components, which is worth mentioning here.}. We obtained the ratio of disc luminosity to corona luminosity ($L_{\rm{disc}}/L_{\rm{coro}}$) at the last moment of each case, which shows that all four cases are in the disc-dominated phase (see the fifth row in Table~\ref{tab:DiffCases}). It should be noted that the obtained disc luminosities are only lower limits for the cases, where the disc extends to the inner boundary set at $r = 10r_{\rm{s}}$ in our model, since the disc may continue to extend beyond our boundary, forming a hot and luminous inner disc. The sharp increase in disc luminosity, shown in Fig.~\ref{fig11} after the inclusion of the physical mechanism (at $\sim 2.45 \times 10^5 r_{\rm{s}}/c$), is likely due to the viscous heating, which causes the internal energy to be predominantly dissipated within the disc.

Fig.~\ref{fig11} shows that a higher accretion rate or viscosity would result in a higher luminosity from both the disc and the corona generally. However, there are subtle differences in their effects. For the disc luminosity, the difference resulting from the accretion rate is obviously larger than that caused by viscosity (see the upper panel of Fig.~\ref{fig11}). In terms of the corona luminosity, the effect of the accretion rate or viscosity does not exhibit a clear clustering similar to that for the disc luminosity, but the distributed order of four curves is the same as that in Fig.~\ref{fig7}, which implies that there may be a positive correlation between the corona luminosity and the evaporation intensity of the accretion disc. This is consistent with physical intuition.

In addition, with the truncation radii and luminosities (approximated as disc luminosities) for each case, we followed the approach of \cite{Nemmen2024} to compare our results with observations of GX 339-4 and other simulations, as is shown in Fig.~\ref{fig12}. Due to the limitation in our model settings, our truncation radii are overestimated, while the luminosities are underestimated. Therefore, we have marked our four data points with downward and rightward arrows in Fig.~\ref{fig12}. It is interesting that the observational data points from different works exhibit the characteristic of clustering into the upper and lower sequences in this figure. The simulation data points of \cite{Nemmen2024}, \cite{Liska2022}, and \cite{Takahashi2016} all fall into the upper sequence, while our points appear between the two sequences, although they may be more likely to be closer to the lower sequence. This may result from the different physical properties between our and theirs (evaporation or condensation). For example, the simulation results of corona condensation such as \cite{Nemmen2024} tend to match the upper sequence, while the ones of disc evaporation proposed in this work tend to approach the lower sequence. It is worth noting that none of our simulations have yet achieved the equilibrium of inflow and outflow, so our results in Fig.~\ref{fig12} are only instantaneous state points where the accretion state evolves from the lower sequence to the upper sequence due to the continuous evaporation of the accretion disc. This evaporation process may be very quick so that a gap appears between these two observational sequences, but our instantaneous numerical points coincidentally appear in this gap.

\subsubsection{Estimation of the $y$ parameter}
\begin{figure}
	\includegraphics[width=\columnwidth]{fig13.png}
	\caption{Profiles of the $y$ parameter for the four cases at two special moments. The horizontal dashed black line represents $y=0.7$, which is the lower criterion that the continuum can be significantly affected by the inverse Compton process.}
	\label{fig13}
\end{figure}
Finally, we focus on an important effect, Compton cooling, which has been ignored in our model to simplify the cooling calculation from the dynamic disc illuminating. According to the studies of DEM \citep[e.g.][]{Liu2002}, Compton cooling in the corona of the inner disc is efficient for luminous sources. Therefore, it is necessary to estimate the $y$ parameter of our model to assess the impact of this assumption. We assume that the opacity of the corona is dominated by scattering. Therefore, the $y$ parameter was obtained by calculating the following integral from zero to a specific optical depth, $\tau_0$, along the sight-line perpendicular to the disc equatorial plane,
\begin{gather}\label{Eq:y-parameter}
  y=\int^{\tau_0}_0\frac{4kT}{m_{\rm{e}}c^2}\max\left(\tau,\tau^2\right)d\tau.
\end{gather}
The function $\max(\tau,\tau^2)$ estimates photon scatterings, which scale as $\sim\tau$ for $\tau<1$ and as $\sim\tau^2$ for $\tau>1$, according to the random-walk theory. 

In Fig.~\ref{fig13}, we show the profiles of the $y$ parameter obtained from all cases. We chose two specific moments for each case, including the initial (blue) and final (red) moments. We adopted three different upper integration limits ($\tau_0=1,10$ and $20$, distinguished by solid, dashed, and dash-dotted lines, respectively) to illustrate the influence of the warm corona. We defined $y>0.7$ (horizontal dashed black lines) as the criterion that the continuum can be significantly affected by the inverse Compton process. Overall, in four cases, the $y$ parameters after adding viscosity heating (red lines) are all significantly increased comparing to the initial discs without viscous heating (blue lines). As a prominent feature, the large $y$ (up to the order of $10^2$) enhanced by the warm corona appears in some discrete regions for all cases (dashed and dash-dotted lines). As is mentioned in Sect.~\ref{subsubsec:Acoustic_sho_hea}, our post-processing tends to overestimate the temperature for optically thick gas. This would not amplify the temperature-independent scattering optical depth, but the scattering events would rapidly increase when $\tau>1$. Thus, the $y$ calculated through Eq.~\ref{Eq:y-parameter} may be reliable for an optically thin hot corona but inflated for an optically thick warm corona due to a large number of scattering events under the overestimated temperature. Nevertheless, we suggest that the warm corona should provide a non-negligible enhancement to the $y$ parameter (implying significant Compton cooling contribution), though it is unlikely to reach an unrealistic level of an order $\sim10^2$. Ignoring the unrealistic effect induced by the warm corona, the $y$ parameters of initial discs without viscous heating (solid blue lines) are quite low. This suggests that the observational ultra-soft state -- lacking observable power-law tail in X-ray spectrum (e.g. the red line in Fig.~9 of \cite{Done2007}) -- may stem from insufficient heating mechanisms in accretion discs.

\section{Conclusion and discussion}\label{Sec:Conclusion}

This is our first study on the interaction between the accretion disc and its corona. We have developed a new 2D axisymmetric, time-dependent HD model consisting of a thin accretion disc, a corona, and a vacuum region, which can be viewed as an extension of the well-known DEM \citep{Meyer2000, Liu2022}. In this work, we implement the GVI tracking algorithm introduced by \cite{Subramaniam2018} into Athena++, allowing us to avoid the computational difficulties posed by the presence of vacuum in FVM codes, while preventing spurious mass generation. Through an adiabatic run that excludes dissipative mechanisms except for the intrinsic shock process embedded in any FVM code, we demonstrate the presence of acoustic shock heating, a mechanism widely studied in the solar physics community but less explored in the context of accretion discs. Then we use the final state of this adiabatic run as the initial condition for four full simulations that include turbulent viscosity, thermal conduction, bremsstrahlung corona cooling, and artificial disc cooling.

Since they suffer from the small time-step induced by the viscosity, full simulations are too slow to achieve the inflow-outflow equilibrium within tolerable running time (however, we find that thermal conduction can partially alleviate this viscous time-step limitation). Therefore, we can only observe the short-term dynamic process of disc evaporation from our four full simulations, which are different from the corona condensation simulations of \cite{Nemmen2024}. Nevertheless, from our numerical results, we can still find some interesting facts: that viscosity dominates the intensity of disc evaporation (Sect.~\ref{subsubsec:DiscEvap}), while the accretion rate primarily determines the disc truncation radius (Sect.~\ref{subsubsec:strctrue}) and the disc luminosity (Sect.~\ref{subsubsec:Lum}). There also appears to be a positive correlation between the corona luminosity and the evaporation intensity (Sect.~\ref{subsubsec:Lum}). The warm corona suggested by observations of AGNs \citep{Petrucci2020} and BHXBs \citep{Jin2024} appears in our numerical results, but only as a thin layer sandwiched between the hot corona and the cold disc (Sect.~\ref{subsubsec:warm_coro}).

We also followed \cite{Nemmen2024} in placing our data points in the diagram of truncation radius versus luminosity to compare them with the ones from the GX 339-4 observations and other simulations. There are two observational sequences in this diagram. Our data points likely approach the lower sequence, while the ones from previous simulations tend to match the upper sequence (Sect.~\ref{subsubsec:Lum}), which may result from the different physical properties between our simulation and previous ones (evaporation or condensation).

Finally, with the estimates of $y$ parameters for our four cases, we assessed the impact of excluding Compton cooling, which has been ignored in our model in order to avoid the computational difficulty of dynamic disc illuminating. Our results indicate that viscous heating significantly increases the $y$-parameter value, while the absence of various potential heating mechanisms provides a plausible explanation for the observationally identified ultra-soft state \citep{Done2007}. We also considered the influence of the warm corona on the $y$ parameter. The warm corona results in unrealistically high values of the virtual $y$ parameter due to the overestimation of temperature and a significant increase in random-walk scattering for optically thick gas in our post-processing. Nevertheless, we contend that the tendency of the warm corona to enhance the $y$ parameter locally, thereby promoting Compton cooling, remains credible.

At the end of this paper, it is important to note that our model is an imperfect simple HD Newtonian model, though it provides the possibility for a deeper insight into disc-corona interaction. To further improve our model, at least, new numerical techniques on viscous processes and dynamic disc illuminating inverse Compton radiation are needed in future work.

\begin{acknowledgements}
	This work was supported by the National Key R\&D Program of China (Grant No. 2023YFA1607902), the National Natural Science Foundation of China (Grant No. 12221003 and 12494572), and the Natural Science Foundation of Fujian Province of China (Grant No. 2023J01008).
\end{acknowledgements}

\bibliographystyle{aa}
\bibliography{aa54428-25corr}
\end{document}